\documentclass[prd,aps,superscriptaddress,onecolumn,%
showpacs,preprintnumbers,amsmath,amssymb,nofootinbib]{revtex4-2}
\usepackage{graphicx}
\usepackage{bm}
\usepackage[colorlinks=true, pdfstartview=FitV, citecolor=blue, urlcolor=blue]{hyperref}
\usepackage{slashed}
\usepackage{array}
\usepackage{mathtools}
\usepackage{mathrsfs}
\usepackage{xcolor}

\graphicspath{{./Figures/}}

\newcommand{\diff}{\mathrm{d}}
\newcommand{\p}{\partial}
\newcommand{\ve}{\varepsilon}
\newcommand{\Diff}{{\mathcal{D}}}

\newcommand{\be}{\begin{equation}}      
\newcommand{\ee}{\end{equation}}      
\newcommand{\bea}{\begin{eqnarray}}      
\newcommand{\eea}{\end{eqnarray}}

\newcommand{\tr}{\mathrm{tr}}
\newcommand{\im}{\mathrm{i}}

\newcommand{\calA}{\mathcal{A}}

\newcommand{\rme}{\mathrm{e}}

\usepackage[normalem]{ulem}

\newcommand{\odre}[1]{\mathcal{O}(#1)}

\newcommand{\eue}[1]{\mathrm{e}^{#1}}
\newcommand{\delt}{\partial_\tau}
\newcommand{\delx}{\partial_x}
\newcommand{\dely}{\partial_y}

\begin{document}
\preprint{YITP-21-92}

\title{Sigma-model analysis of $SU(3)$ antiferromagnetic spins on the triangular lattice}

\author{Itsuki Takahashi}
\email{takahashi.itsuki.87n@kyoto-u.jp}
\affiliation{Department of Physics, Kyoto University, 
  Kyoto 606-8502, Japan}

\author{Yuya Tanizaki}
\email{yuya.tanizaki@yukawa.kyoto-u.ac.jp}
\affiliation{Yukawa Institute for Theoretical Physics, Kyoto University, 
  Kyoto 606-8502, Japan}

\date{\today}

\begin{abstract}
Using field-theoretic techniques, we study the $SU(3)$ analogue of anti-ferromagnetic Heisenberg spin model on the triangular lattice putting the $p$-box symmetric representation on each site. 
Taking the large-$p$ limit, we show that the low-energy effective theory is described by a $(2+1)$-dimensional relativistic $SU(3)/U(1)^2$ nonlinear sigma model. 
Since the target space has a nontrivial homotopy $\pi_2(SU(3)/U(1)^2)\simeq \mathbb{Z}^2$, this model has two kinds of magnetic skyrmions, which can be created and annihilated by monopole instantons. 
By careful analysis of the Wess-Zumino term in the spin coherent path integral, we compute the Berry phase for these monopoles and it produces the destructive interference. 
This restricts possible perturbations of the effective Lagrangian by monopole operators, and we see that the valence-bond-solid (VBS) phase should have degenerate ground states when $p\not\in 3\mathbb{Z}$. 
We also compute 't Hooft anomalies to constrain possible phases of this system, and a direct phase transition between N\'eel and VBS phases is supported from the anomaly matching.  
\end{abstract}

%\pacs{}
%\keywords{}

\maketitle

\section{Introduction} \label{sec:intro}

Study of quantum many-body spin systems is an important and interesting subject that attracts not only condensed-matter communities but also high-energy communities. 
Quantum spin systems provide us many examples, in which quantum nature gives a new physics that does not appear in classical analogue. 
The most basic example is the Heisenberg spin system, 
\begin{equation} \label{Heisenberg}
H=2J\sum_{\langle i,i'\rangle} \hat{\bm{S}}(i)\cdot\hat{\bm{S}}(i')\;, 
\end{equation}
where $\hat{\bm{S}}(i)$ is the spin vector at the site $i$. 
When $J>0$, the anti-parallel spins are energetically favored, and the N\'eel order appears at the classical level when the underlying lattice is frustration-free. 

Quantum field theory (QFT) is one of the basic tools to take into account the quantum fluctuations. 
In the context of many-body spin systems, seminal papers by Haldane~\cite{Haldane:1982rj,Haldane:1983ru} have shown that the  antiferromagnetic spin chains can be described by the $(1+1)$-dimensional nonlinear sigma model with the topological $\theta$ term. 
By finding that the size of spin $s$ is related to the topological term as $\theta=2\pi s$, he conjectured that the low-energy dynamics of spin chain is drastically different depeding on whether $s$ is half-integer or integer: For half-integer spins, the spin chain has a gapless excitation, while the system is gapped for integer spins. 
This is a surprising result uncovered by applying QFT to spin systems. 

Antiferromagnetic Heisenberg model on the square lattice gives a prototypical example of deconfined quantum criticality (DQC)~\cite{Senthil:2003eed, PhysRevB.70.144407}, and QFT also plays an important role in that context. 
In classical statistical physics, the low-energy effective theory of the phase transition is usually given by some continuum field theory with the order-parameter field. 
However, this Ginzburg-Landau-Wilson paradigm is insufficient to understand DQC, since the effective degrees of freedom contains emergent gauge fields and the order parameters are fractionalized. 
The low-energy QFT becomes the compact $U(1)$ gauge theory coupled to $2$-flavor complex scalar field, which becomes the $\mathbb{C}P^1$ sigma model in the N\'eel order phase, while the monopoles are relevant in the valence bond solid (VBS) phase. 
In each phase, we can use confined, or gauge-singlet, degrees of freedom for the effective description, but the treatment of deconfined degrees of freedom is necessary to understand the N\'eel-VBS transition. 
In the actual realization of the N\'eel-VBS transition, the higher-order spin interaction should be added as the driving force of this transition~\cite{Sandvik:2006fpf}, and many observations obtained from QFT turn out to be robust for such extensions. 

In this paper, we consider the $SU(3)$ antiferromagnetic spins on the triangular lattice, and generalize the N\'eel-VBS transition to $SU(3)$ spin systems. 
Such a generalization from $SU(2)$ spins to $SU(N)$ spins has been recently considered in the context of spin chains~\cite{Bykov:2011ai, Bykov:2012am, Lajko:2017wif, Tanizaki:2018xto, Ohmori:2018qza, Hongo:2018rpy, Wamer:2019oge, Affleck:2021ypq}, which leads to the $SU(N)$ generalization of the Haldane conjecture. 
On the triangular lattice, the classical $SU(3)$ spins can be set to the N\'eel order without any frustrations thanks to the tripartite lattice structure. 
For the fundamental $SU(3)$ representation, this system is equivalent to the $SU(3)$-symmetric point of the $SU(2)$ spin-$1$ bilinear-biquadratic (BBQ) model~\cite{tsunetsugu2006}. 
We also note that $SU(N)$ spin systems can be experimentally realized by ultracold atoms with alkaline earth metals~\cite{ PhysRevLett.92.170403, 1367-2630-11-10-103033, gorshkov2010two, taie20126, Zhang1467, 0034-4885-77-12-124401, CAPPONI201650}, which would give us an opportunity to verify theoretical predictions about these generalized spin systems in a near future. 

We first derive the effective Lagrangian using the spin coherent path integral in the N\'eel ordered phase, and obtain the nonlinear sigma model with the flag-manifold target space, $SU(3)/U(1)^2$. 
This result is consistent with previous works by Smerald and Shannon~\cite{PhysRevB.88.184430}, and we here present more detailed derivation. 
In this paper, we consider the $p$-box symmetric representation for each spin, and we take the Haldane limit, $p\to \infty$, in order to derive the effective field theory. 
We obtain the $SU(3)/U(1)^2$ sigma model without any topological terms in the end, but it turns out that careful examination of the Wess-Zumino term is required to reach this conclusion. 
Even though we expect that the $SU(3)/U(1)^2$ sigma model is a valid description even for small $p$'s in view of symmetry, the large-$p$ limit is taken for rigorous semiclassical derivation from the microscopic Hamiltonian.

Since $\pi_2(SU(3)/U(1)^2)\simeq \mathbb{Z}^2$, there are two kinds of magnetic skyrmions in the low-energy effective theory. 
Within the continuum approximation, the numbers of skyrmions are conserved quantities, but there is no corresponding $U(1)\times U(1)$ symmetry in the lattice model. 
Therefore, there should be some high-energy process that changes the skyrmion numbers, which can be understood as the effect of dynamical monopoles in the effective field theory. 
As these monopoles introduces singularities in the continuum description, the Wess-Zumino term may introduces some phase factor in the path integral for such processes as in the case of $SU(2)$ spins~\cite{Haldane:1988zz, Read:1989zz, Read:1989jy, Read:1990zza}. 
Computing the Berry phase for the monopole events, we find that such monopole events have the descructive interference when $p\not=0 \bmod 3$, 
and we conclude that the $\mathbb{Z}_3\times \mathbb{Z}_3$ subgroup of the emergent $U(1)\times U(1)$ symmetry comes out of the underlying lattice symmetry, so it can be regarded as an exact symmetry of the system. 
Using this information, we compute the 't~Hooft anomalies of this system, and constrain its possible low-energy dynamics by the anomaly matching condition. 
Let us emphasize that the result obtained from symmetry and 't~Hooft anomaly is rigorous, and it is robust for any local and symmetric deformations of the lattice Hamiltonian. 
Combined with the consideration on dynamics, this gives the strong evidence for the direct phase transition between the N\'eel and VBS orders.

This paper is organized as follows. 
In Sec.~\ref{sec:NLS}, we explain our derivation of the effective theory of $SU(3)$ antiferromagnetic spins on the triangular lattice. 
In Sec.~\ref{sec:symmetry}, we compare the symmetries between the lattice model and the $SU(3)/U(1)^2$ nonlinear sigma model. 
In Sec.~\ref{sec:tunnel}, we construct various skyrmion configuration and associated tunneling processes. We then compute the Berry phase of the monopole amplitudes, and give the graphical rule to compute the Berry phase conveniently. 
In Sec.~\ref{sec:phase}, we examine 't~Hooft anomalies of this theory, and argue that there should be a direct phase transition between the N\'eel and VBS phases. Section~\ref{sec:conclusion} is devoted to the summary and the discussion for possible directions of future studies. 
We include appendices to give detailed computations and discussions. 

\section{Non-linear Sigma Model}  \label{sec:NLS}

In this section, we describe the low-energy effective description of $SU(3)$ anti-ferromagnetic spin systems on triangular lattice. 
It turns out that the low-energy effective theory can be described by the relativistic non-linear sigma model, and the sigma-model target space is given by the flag manifold $SU(3)/U(1)^2$. 

\subsection{Lattice $SU(N)$ Heisenberg model}  \label{ssec:model}

In this subsection, we describe the lattice Hamiltonian for $SU(N)$ Heisenberg model. 
In this model, $SU(N)$ spins are put on lattice sites, and they interact with the nearest neighbor couplings.  
The Hamiltonian is given by
\begin{equation} \label{hamiltonian}
H=J\sum_{\langle i,i'\rangle}\sum_{v,w=1}^{N} \hat{S}_{vw}(i)\hat{S}_{wv}(i')+\cdots \;,
\end{equation}
where $J>0$ denotes the anti-ferromagnetic Heisenberg coupling, and the sum runs over the nearest neighbor pairs of sites $\langle i, i'\rangle$. 
The ellipsis $\cdots$ represents possible higher-order spin interactions that are short-ranged and have the same symmetry with the leading one. 
Here, $\hat{S}_{vw}$ is an $SU(N)$ spin operator with the indices $v,w=1,\ldots,N$, which satisfies the following commutation relation 
\begin{equation} \label{spinalgebra}
[\hat{S}_{vv'},\,\hat{S}_{ww'}]=\delta_{vw'}\hat{S}_{v'w}-\delta_{v'w}\hat{S}_{vw'} \;. 
\end{equation}
%Repeated greek subscripts are summed over 1 to $N$. 
In this paper, we concentrate on totally symmetric spin representations. 
Totally symmetric representations can be labelled by $p\in\mathbb{N}$, as they have Young tableaux with 1 row and $p$ columns, and we call it the $p$-box symmetric representation. 
When $N=2$, the leading term of the Hamiltonian~\eqref{hamiltonian} reduces to the familiar form of Heisenberg model~\eqref{Heisenberg}, where the magnitude of the spin $s$ is given by $2s=p$. 
Hence the model~\eqref{hamiltonian} is an $SU(N)$ extension of Heisenberg model~\eqref{Heisenberg}. 

In order to study the ground-state property of the Hamiltonian \eqref{hamiltonian}, we compute the partition function $Z=\mathrm{Tr}\,\mathrm{e}^{-\beta H}$ using the imaginary-time path integral. 
For this purpose, we introduce the Schwinger boson~\cite{PhysRevB.38.316} as a convenient description of the spin coherent states. 
We decompose the $SU(N)$ spin operator as 
\begin{equation} \label{schwingerboson1}
\hat{S}_{vw}(i)=\hat{a}_v^\dagger(i) \hat{a}_w(i) \;,
\end{equation}
where $\hat{a}^\dagger_v$ and $\hat{a}_v$ are creation and annihilation operators of bosonic harmonic oscillators,
\begin{equation} \label{bosoncom}
\bigl[\hat{a}_v(i),\, \hat{a}_w^\dagger(i')\bigr]=\delta_{vw}\delta_{ii'} \;,\;\;
\bigl[\hat{a}_v(i),\, \hat{a}_w(i')\bigr]=
\bigl[\hat{a}_v^\dagger(i),\, \hat{a}_w^\dagger(i')\bigr]=0 \; .
\end{equation}
It is straightforward to confirm that $\hat{S}_{vw}$ defined by \eqref{schwingerboson1} satisfies the same commutation relation~\eqref{spinalgebra}. 
The $p$-box symmetric representation can be obtained by imposing the constraint on the occupation number as follows:
\begin{equation} \label{schwingerboson2}
\sum_v \hat{a}_v^\dagger(i)\hat{a}_v(i)=p \;. 
\end{equation}
We introduce the $SU(N)$ spin coherent state by  
\begin{equation} \label{coherentdif}
|\vec{\Phi}(i)\rangle=\frac{1}{\sqrt{p!}}\left[\sum_{v}\Phi_v(i) \hat{a}_v^\dagger(i)\right]^p|0\rangle \;, 
\end{equation}
where $\vec{\Phi}(i)=(\Phi_v(i))_{v=1,\ldots,N}\in\mathbb{C}^N$ with $|\vec{\Phi}|^2=1$, and $\vec{\Phi}$ is in the fundamental representation under the $SU(N)$ spin rotation. 
Notably, $|\vec{\Phi}\rangle$ satisfies the constraint~\eqref{schwingerboson2}, so it gives a vector in the $p$-box symmetric representation.  
Using the following properties of the coherent state $|\vec{\Phi}\rangle$ (see Appendix~\ref{ssec:spin_coherent} for their derivation),
\begin{align}
\langle\vec{\Phi}|\vec{\Phi}\rangle=1 &\qquad \text{(normalization)}\;, \label{coherent1} \\
\int \diff \Omega_{\Phi}\,|\vec{\Phi}\rangle\langle\vec{\Phi}|=1 &\qquad \text{(completeness)}\;, \label{coherent2} \\
\langle\vec{\Phi}'|\vec{\Phi}\rangle=(\vec{\Phi}'^\ast\cdot\vec{\Phi})^p &\qquad \text{(overlap)}\;, \label{coherent3} \\
\langle\vec{\Phi}|\hat{S}_{vw}|\vec{\Phi}\rangle=p\,\Phi_v^\ast\Phi_w &\qquad \text{(spin operator)}\;, \label{coherent4} 
\end{align}
we obtain the path-integral expression of the partition function,
\begin{equation}  \label{partition}
Z = \int \mathcal{D}\Phi \,\exp(-S[\Phi]) \;,
\end{equation}
where the Euclidean action is given by 
\begin{equation} \label{action} 
S = \int_0^\beta \diff \tau\left\{
p\sum_i\left(\vec{\Phi}^\ast(i,\tau)\cdot\partial_\tau\vec{\Phi}(i,\tau)\right)+Jp^2\sum_{\langle i,i'\rangle}\left|\vec{\Phi}^\ast(i,\tau)\cdot\vec{\Phi}(i',\tau)\right|^2
\right\} .  
\end{equation}
Here, the path-integral measure $\Diff\Phi$ is given by the product of $SU(N)$-invariant measure $\diff \Omega_\Phi\propto \diff^N \vec{\Phi}^*\diff^N \vec{\Phi} \delta(|\vec{\Phi}|^2-1)$ over the spacetime points. 
The first term of Eq.~\eqref{action} is called as the Wess-Zumino term or as the Berry phase term, which we will denote as $S_\mathrm{B}$, and the second term is the exchange term coming from the Hamiltonian, which describes the classical energy of each  configuration. 
As we will begin with the sigma-model analysis of the N\'eel phase, we have only taken into account the leading term of the Hamiltonian~\eqref{hamiltonian} in \eqref{action}. 
Note that the Wess-Zumino term $S_\mathrm{B}$ is independent of the coupling constant $J$. 

Let us also point out that the action~\eqref{action} is $U(1)$ gauge invariant under the local transformation, $\vec{\Phi}(i,\tau)\mapsto \rme^{\im \vartheta(i,\tau)}\vec{\Phi}(i,\tau)$, which comes out of the decomposition of spin operators~\eqref{schwingerboson1}.  
The gauge invariance of the second term is obvious, but the Wess-Zumino term changes as $\vec{\Phi}^\ast(i,\tau)\cdot\partial_\tau\vec{\Phi}(i,\tau)\mapsto \vec{\Phi}^\ast(i,\tau)\cdot\partial_\tau\vec{\Phi}(i,\tau)+\im\, \partial_\tau \vartheta(i,\tau)$. 
Because of the $2\pi$ periodicity of the gauge parameter, $\vartheta(i,\tau+\beta)=\vartheta(i,\tau) \bmod 2\pi$, its imaginary-time integral changes only by $2\pi \im\, p \mathbb{Z}$, so the path-integral weight is gauge invariant for integer $p$. 

\subsection{Effective Lagrangian}  \label{ssec:efflag}

\begin{figure} \centering
\includegraphics[scale=0.35]{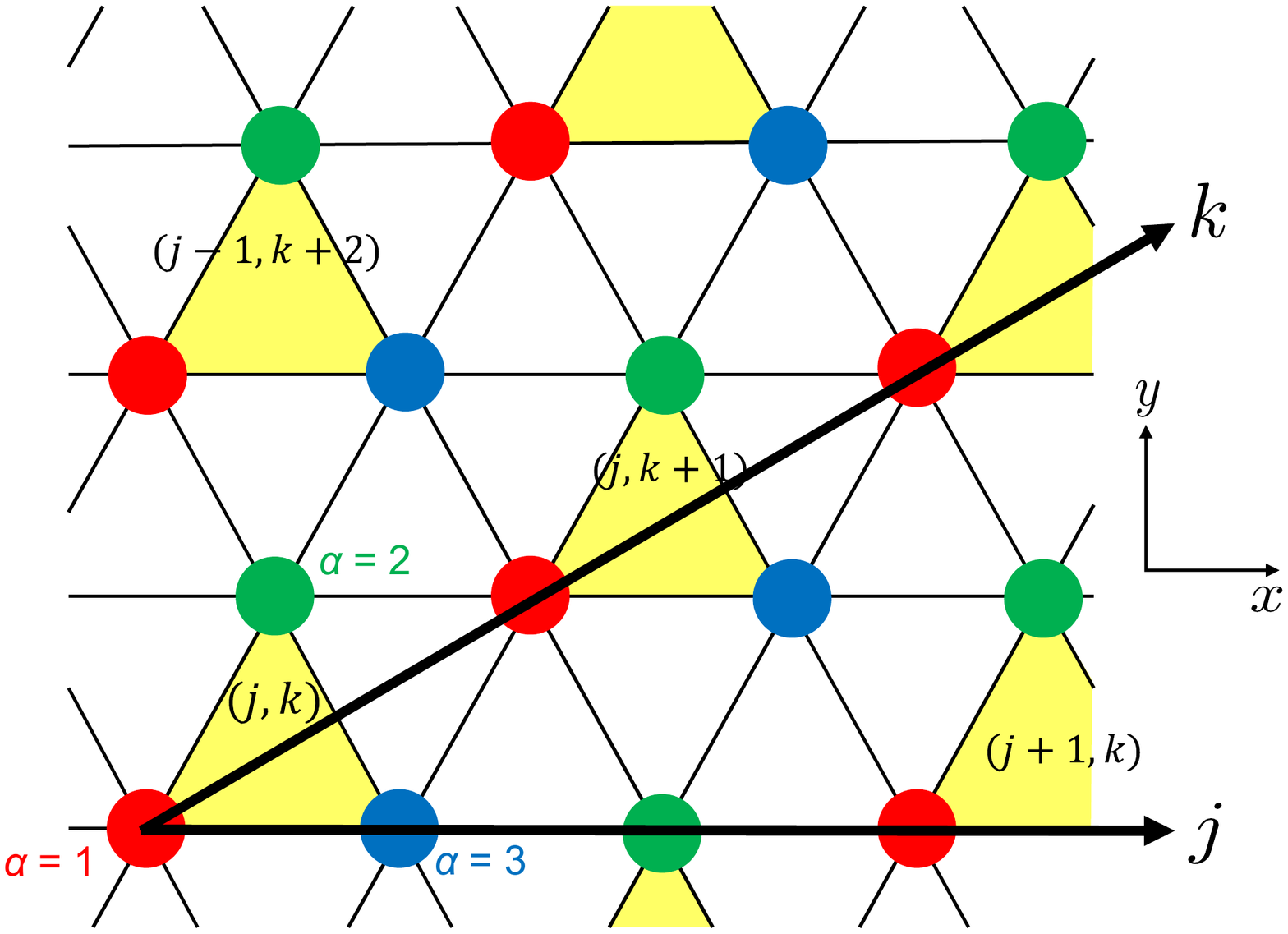}
\caption{Illustration of the N\'eel ordered state: On the triangular lattice, $SU(3)$ spins does not have frustrations. 
To find the continuum field theory, we combine three $SU(3)$ spins in each yellow unit cell into a single field, and the location of those unit cells is designated by $(j,k)\in \mathbb{Z}^2$.
We denote the sublattices shown with red, green, and blue by $\alpha=$ 1, 2 and 3, respectively, so the $SU(3)$ spins $\Phi(i)$ can be written as $\Phi_{\alpha}(j,k)$. }
\label{fig:SU(3)coordinate}
\end{figure}

Let us now restrict our attention to the case of $SU(3)$ spins on the triangular lattice. 
The classical ground state is given by the N\'eel ordered state as drawn in Fig.~\ref{fig:SU(3)coordinate}: In order to minimize the classical energy, it is sufficient to achieve $\vec{\Phi}^*(i)\cdot \vec{\Phi}(i')=0$ for the nearest neighboring states. 
Because of the tripartite lattice structure,  we can assign, e.g., 
\begin{equation}
    \vec{\Phi}=\begin{pmatrix}
    1\\0\\0
    \end{pmatrix},\,\,
    \begin{pmatrix}
    0\\1\\0
    \end{pmatrix},\,\, \mbox{and}\,\,
    \begin{pmatrix}
    0\\0\\1
    \end{pmatrix},
\end{equation}
for each sublattice denoted by red, green, and blue blobs in Fig.~\ref{fig:SU(3)coordinate}, respectively, to satisfy that condition. 
In the limit $p\to \infty$, the classical N\'eel state becomes more and more accurate description of the true ground state, and we start taking into account the quantum fluctuations at finite $p$.

Before the derivation of the effective Lagrangian, let us guess what kinds of QFT would appear as the low-energy effective theory. 
We note that the classical N\'eel state spontaneously breaks the $SU(3)$ spin rotational symmetry to its maximal Abelian subgroup, 
\begin{equation}
    SU(3)\to U(1)\times U(1)\;. 
\end{equation}
Therefore, there are six Nambu-Goldstone (NG) bosons with the linear dispersion relation at the classical level. 
Moreover, as we can see in Fig.~\ref{fig:SU(3)coordinate}, the classical configuration has $\mathbb{Z}_3$ rotational symmetry, so the propagation along $x$ and $y$ directions must have the same speed for low momenta. 
According to the standard technique of the nonlinear realization~\cite{Coleman:1969sm}, we can expect that the effective theory is given by the relativistic nonlinear sigma model with the target space $SU(3)/U(1)^2$. 

We here explain the outline to obtain the $SU(3)/U(1)^2$  sigma model using the semiclassical approximation around the classical N\'eel order in the large-$p$ limit, and more detailed derivation can be found in Appendix~\ref{sec:A}. 
To obtain the low-energy effective theory, 
we introduce the unit cell as shown with the yellow triangle in Fig.~\ref{fig:SU(3)coordinate}. 
Each unit cell is designated by the label $(j,k)\in \mathbb{Z}^2$, which is related to $x,y$ coordinates as $(x,y)=a(3j+ \frac{3}{2}k ,\frac{\sqrt{3}}{2}k)$ with the lattice constant $a$. 
Each sites inside the unit cell is denoted by $\alpha=1,2,3$, which corresponds to the red, green, blue sites in Fig.~\ref{fig:SU(3)coordinate}, respectively, and we can now rewrite the label $i$ of each site by the triple $(j,k,\alpha)$: 
\begin{equation}
\vec{\Phi}_\alpha(j,k,\tau):=\vec{\Phi}(i,\tau)\;. 
\end{equation}
We now separate the low-energy fluctuations, which are gapless at the classical level, from the high-energy fluctuations, which describes the fast fluctuations inside the unit cell.  
Following the proposal of Ref.~\cite{Lajko:2017wif}, we decompose the $SU(3)$ spins inside the unit cell at $(j,k)$ as 
\begin{equation}  \label{SU(3)fluc}
\left( \begin{array}{c}
\vec{\Phi}_1(j,k)^\mathrm{T} \\ \vec{\Phi}_2(j,k)^\mathrm{T} \\ \vec{\Phi}_3(j,k)^\mathrm{T}
\end{array} \right) =
L(j,k)\cdot U(j,k) \;, 
\end{equation}
where $L(j,k)(=L(j,k)^\dagger)$ describes the fast modes,
\begin{equation}  \label{SU(3)matrixL}
L(j,k)= \left( \begin{array}{ccc}
\!\!\!\sqrt{1-\tfrac{a^2}{p^2}(|L_{12}|^2+|L_{13}|^2)}\!\!\!  & \tfrac{a}{p}L_{12} & \tfrac{a}{p}L_{13} \\
\tfrac{a}{p}L_{12}^\ast & \!\!\!\sqrt{1-\tfrac{a^2}{p^2}(|L_{12}|^2+|L_{23}|^2)}\!\!\!  & \tfrac{a}{p}L_{23} \\
\tfrac{a}{p}L_{13}^\ast &\tfrac{a}{p}L_{23}^\ast & \!\!\!\sqrt{1-\tfrac{a^2}{p^2}(|L_{13}|^2+|L_{23}|^2)}\!\!\!
\end{array} \right) 
\end{equation}
and $U(j,k)$ denotes the slow modes and lives in the classical vacua, $U(j,k)\in U(3)$: Taking the parametrization of $U(j,k)$ as 
\begin{equation}  \label{SU(3)matrixU}
U(j,k) = \left( 
\begin{array}{c}
\vec{\phi}_1(j,k)^\mathrm{T} \\ \vec{\phi}_2(j,k)^\mathrm{T} \\ \vec{\phi}_3(j,k)^\mathrm{T}
\end{array} 
\right),
\end{equation}
we obtain the orthonormality constraint
\begin{equation}  \label{orthonomal}
\vec{\phi}_\alpha^\ast(j,k)\cdot\vec{\phi}_\beta(j,k) = \delta_{\alpha\beta} \; . 
\end{equation}
Let us check if the number of degrees of freedom matches before and after this decomposition. On the left-hand-side of \eqref{SU(3)fluc}, there are $15$ degrees of freedom, $3$ of which are unphysical due to $U(1)^3$ gauge redundancy. 
On the right-hand-side, $U(j,k)$ has $9$ degrees of freedom, $3$ of which are again unphysical by gauge redundancy, and this is consistent with the existence of $6$ massless NG bosons at the classical level. 
Since $L(j,k)$ has $6$ degrees of freedom, the number of degrees of freedom matches on both sides. We also note that the form of $L(j,k)$ is chosen so that $|\vec{\Phi}|^2=1$ is satisfied. 
Using one of the $U(1)$ gauge redundancies, we can impose that 
\begin{equation}\label{eq:gauge_fixing}
    \det U(j,k)=1\quad (\Leftrightarrow \, \vec{\phi}_{1}\cdot(\vec{\phi}_{2}\times \vec{\phi}_{3})=1)\;, 
\end{equation}
so we regard $U(j,k)\in SU(3)$ in the following.

Substituting the decomposition~\eqref{SU(3)fluc} into the Euclidean action~\eqref{action}, we find that $L(j,k)$ describes the massive excitations at the lattice scale, and we can integrate it out by the Gaussian integration (see Appendix~\ref{ssec:apSU(3)NLS} for details). 
The effective action is given by~\cite{PhysRevB.88.184430, Pimenov:2017nzi}
\begin{equation} \label{SU(3)effaction} 
S_\mathrm{eff} = \int\diff^3 x \:
\frac{1}{g_{\mathrm{eff}}}\sum_{\alpha=1}^3\left( \left|\partial_\mu\vec{\phi}_\alpha\right|^2-\left|\vec{\phi}_\alpha^\ast\cdot\partial_\mu\vec{\phi}_\alpha\right|^2 \right)  \;.
\end{equation}
Here, we defined the effective coupling constant $g_{\mathrm{eff}}=\frac{3\sqrt{3}}{\sqrt{2}p}a$, which becomes small in the large-$p$ limit. 
To obtain the relativistic action~\eqref{SU(3)effaction}, we rescaled the imaginary time as $\tau\mapsto v \tau$, where $v=\frac{3 p}{\sqrt{2}}Ja$ is the speed of gapless excitation.  
This is the $(2+1)$-dimensional $SU(3)/U(1)^2$ nonlinear sigma model, which describes the quantum fluctuations around the classical N\'eel ordered state. 

For the later purpose, it is convenient to rewrite the $SU(3)/U(1)^2$ sigma model into the $SU(3)$ sigma model coupled to the $U(1)\times U(1)$ gauge theory. 
We introduce the auxiliary $U(1)$ gauge fields $a_{\alpha}=a_{\alpha,\mu}\diff x^\mu$ by 
\begin{equation}  \label{SU(3)gaugeba}
a_{\alpha}=\im\, \vec{\phi}_\alpha^\ast\cdot\diff \vec{\phi}_\alpha \;, 
\end{equation}
and the condition~\eqref{eq:gauge_fixing} gives the following constraint,
\begin{equation}\label{eq:constraint_gaugefield}
    \sum_{\alpha=1}^{3}a_\alpha=0\;. 
\end{equation}
The effective action~\eqref{SU(3)effaction} can be written as 
\begin{equation}  \label{SU(3)effactionA} 
S_\mathrm{eff} = \int\diff^3 x\,
\frac{1}{g_{\mathrm{eff}}}\sum_{\alpha=1}^3\left|(\partial_\mu+\im\, a_{\alpha,\mu})\vec{\phi}_\alpha\right|^2 \;. 
\end{equation}
The $U(1)\times U(1)$ gauge invariance of $S_{\mathrm{eff}}$ becomes more manifest in this form. 
We can now regard that $a_{\alpha}$ are $U(1)$ gauge fields, which are independent of $\phi_{\alpha}$. 
Since the Lagrangian is quadratic in $a_{\alpha,\mu}$, the path integral can be performed exactly by solving its classical equation of motion, which is nothing but \eqref{SU(3)gaugeba}. Therefore, both viewpoints give equivalent descriptions.

\subsection{Absence of topological terms}  \label{ssec:Hopf}

In the previous section, we find that the effective action $S_\mathrm{eff}$ can be obtained from the lattice Hamiltonian using the semiclassical approximation for large $p$, following Ref.~\cite{Lajko:2017wif} that computes the effective theory of $(1+1)$d $SU(3)$ spin chains. 
We here have to emphasize that any topological terms do not appear in the effective Lagrangian, according to our computations given in Appendix~\ref{ssec:apSU(3)NLS} and \ref{ssec:apSU(3)top}. 
Since this is a subtle but important feature of the low-energy effective theory, let us summarize how the topological terms vanish in our derivation. 

Let us first review the logic given by Haldane~\cite{Haldane:1988zz}, which shows the absence of the topological term for the $SU(2)$ antiferromagnets on the square lattice. 
The $2$d square lattice can be regarded as a stack of $1$d chains, and each $1$d spin chain can be described by the $2$d $\mathbb{C}P^1$ sigma model with $\theta=2\pi s$. By combining two $1$d chains together, the total $\theta$ angle is equivalent to $0$ up to its $2\pi$ periodicity just by the assumption of the continuity of fields. 
In this way, Haldane showed that the topological term does not appear from the Wess-Zumino term of the spin coherent path integral in any spatial dimensions $>1$. 

We can use the similar logic to claim the absence of the topological terms in $SU(3)$ antiferromagnets on the triangular lattice. 
As shown in Refs.~\cite{Bykov:2011ai, Bykov:2012am, Lajko:2017wif}, $1$d $SU(3)$ antiferromagnetic spin chain is described by the $2$d $SU(3)/U(1)^2$ sigma model with $\theta_1={2\pi \over 3}p$ and $\theta_2={4\pi\over 3}p$. 
When regarding $2$d triangular lattice as a stack of $1$d chains, we combine three consecutive $1$d chains into an effective single chain, then both $\theta$ angles become $0$ up to its $2\pi$ periodicity. 
This means that, when the number of sites is a multiple of $3$ for each direction so that we can put the spatical periodic boundary condition, the topological term should be absent for the effective $3$d $SU(3)/U(1)^2$ sigma model. 

We have checked this expectation explicitly in Appendix~\ref{ssec:apSU(3)NLS} and \ref{ssec:apSU(3)top}. 
Naive derivation of the effective action has the following topological-like term
\begin{equation}  \label{SU(3)efftop} 
\frac{2\pi \im }{3a}p\left[ 
\frac{1}{\sqrt{3}}\int \diff y\,Q_1^{x\tau}(y)-\int \diff x\,Q_2^{y\tau}(x)-\frac{1}{\sqrt{3}}\int \diff y\,Q_3^{x\tau}(y)
\right] , 
\end{equation}
in addition to the kinetic term~\eqref{SU(3)effaction}. 
Here, $Q_{\alpha}^{\mu\nu}$ denotes the topological charge of $a_{\alpha}$ along the $\mu\nu$ directions,
\begin{equation}   \label{Qdef} 
Q_\alpha^{\mu\nu}={1\over 2\pi}\int_{\mathbb{R}^2_{\mu\nu}} \diff a_{\alpha} \in \mathbb{Z} \;. 
\end{equation}
Replacing the spatial integration in \eqref{SU(3)efftop} by the summation over lattice sites after suitable change of coordinates, we find that \eqref{SU(3)efftop} vanishes mod $2\pi \im $ when we can compactify the spatial lattice into $T^2$ with the periodic boundary condition.
We therefore conclude that $S_{\mathrm{eff}}$ does not contain topological terms. 

Although topological terms do no appear in our specific lattice model, it is still an interesting question to ask what kinds of topological terms can be added to the $SU(3)/U(1)^2$ sigma model. 
Since $\pi_3(SU(3)/U(1)^2)\simeq \mathbb{Z}$, one may wonder if we can have a $2\pi$ periodic $\theta$ angle using the Hopf-like invariant. 
Explicit construction of the Hopf-like invariant for the flag manifold can be found in Refs.~\cite{Kisielowski:2013ina,Amari:2018gbq}. 
However, it has been shown that such a Hopf-like term cannot appear in the Lagrangian of the relativistic $SU(3)/U(1)^2$ sigma model, as long as it satifies the unitarity and locality axioms of QFT~\cite{Kobayashi:2021qfj} (see also Refs.~\cite{Freed:2016rqq, Freed:2017rlk, Yonekura:2018ufj}). 
This is because the possible topological terms should be characterized by the bordism group, $\Omega^{\mathrm{spin/SO}}_*(SU(3)/U(1)^2)$, instead of the homotopy, and we cannot find corresponding elements for the $\theta$ angles in $\Omega^{\mathrm{spin/SO}}_{3,4}(SU(3)/U(1)^2)$ including discrete and continuous ones. 
Ref.~\cite{Kobayashi:2021qfj} has shown that there are two Chern-Simons terms as possible topological terms, but it is also shown that they cannot appear without explicit breaking of the underlying lattice symmetry. 

\section{Symmetries}  \label{sec:symmetry}

In this section, we discuss the symmetry property of the lattice $SU(3)$ Heisenberg model on the triangular lattice, and its translation to the $SU(3)/U(1)^2$ sigma model description. 
Since the symmetry is an exact property of the system, the low-energy QFT should have the same symmetry properties. 
We also discuss the emergent symmetry that appears in the low-energy QFT but is absent in the lattice model. 

The lattice Hamiltonian~\eqref{hamiltonian} is invariant under the $SU(3)$ spin rotation, which acts on the spin operator $\hat{S}_{\alpha\beta}$ in the adjoint representation, 
\begin{equation}
    \hat{S}\to V \hat{S} V^{\dagger}\;,
\end{equation}
with $V\in SU(3)$. Since the center subgroup $\mathbb{Z}_3\subset SU(3)$ acts as the trivial transformation, the spin rotations with the faithful action is given by the quotient group, $PSU(3)=SU(3)/\mathbb{Z}_3$. 

\begin{figure} \centering
\includegraphics[scale=0.30]{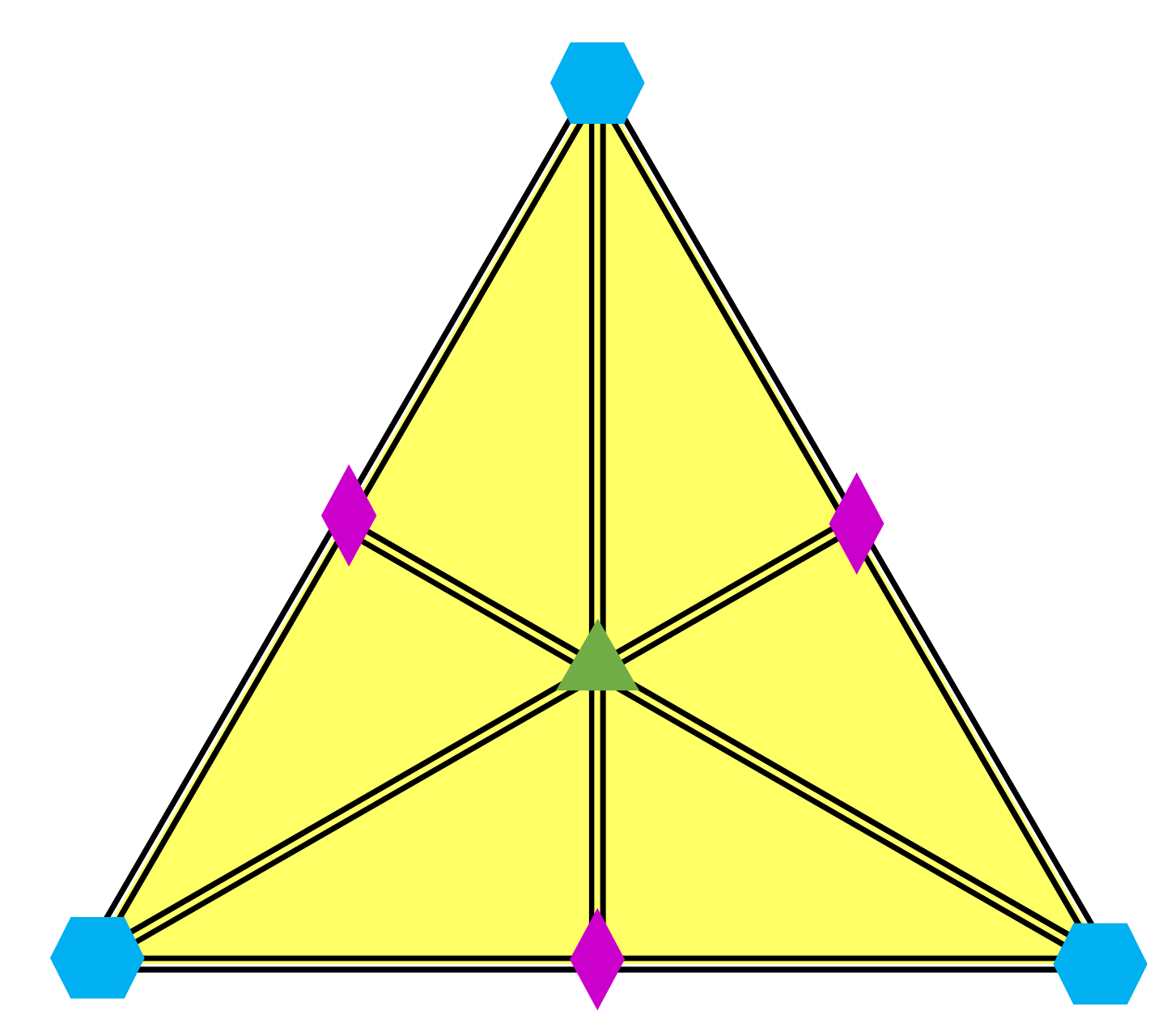}
\caption{Fundamental structure of the wallpaper group, $p6m$. The symbols, diamond, triangle, and hexagon, denote the center of $\pi$, $2\pi/3$, and $2\pi/6$ rotations, respectively. Double line denotes axis of reflection. }
\label{fig:p6m}
\end{figure}

Since we assign the same representations on every site of the triangular lattice, the system has the isometry symmetry  of the triangular lattice. 
It is called as the wallpaper group, $p6m$, in the crystallographic notation, which is explained in Fig.~\ref{fig:p6m}. 
Here, we draw the primitive cell of the triangular lattice, and express the rotation centers of degree $2$, $3$, and $6$ by diamonds, triangles, and hexagons, respectively. 
There are also mirror isometries, whose reflection centers are denoted by the double lines. 
It contains the dihedral group $D6$ as a point group, which is generated by the $2\pi/6$ rotation around the site and one of the reflection. 

These spin and lattice symmetries should be inherited by the $SU(3)/U(1)^2$ sigma model. 
Identification of those symmetry realizations, we can understand the possible perturbations in the effective Lagrangian away from the semiclassical limit of the N\'eel phase. 
Continuum Lagrangian should be invariant under the following transformations, $\phi\mapsto \phi'$:

\begin{enumerate}
\item $U(1)^2$ gauge symmetry, due to the use of Schwinger bosons~\eqref{schwingerboson1},
\begin{equation}  \label{SU(3)symG}
\vec{\phi}'_\alpha(x,y,\tau)=\mathrm{e}^{i\vartheta_\alpha(x,y,\tau)}\vec{\phi}_\alpha(x,y,\tau) \;,\quad \sum_{\alpha=1}^{3} \vartheta_\alpha=0 \;.
\end{equation}
\item $PSU(3)=SU(3)/\mathbb{Z}_3$ symmetry, 
\begin{equation}  \label{SU(3)symS}
\vec{\phi}'_\alpha(x,y,\tau)=V\vec{\phi}_\alpha(x,y,\tau) \;,\quad V\in SU(3) \;.
\end{equation}
\item $(\mathbb{Z}_3)_{\mathrm{trans}}$ cyclic symmetry, originated from the $1$ unit lattice translation,
\begin{equation}  \label{SU(3)symZ}
\bigl[\, \vec{\phi}'_1, \; \vec{\phi}'_2, \; \vec{\phi}'_3 \,\bigr](x,y,\tau)=\bigl[\, \vec{\phi}_2, \; \vec{\phi}_3, \; \vec{\phi}_1 \,\bigr](x,y,\tau) \;.
\end{equation}
\item $x$-axis reflection symmetry (for the axis through $\alpha=1$ vertices),
\begin{equation}  \label{SU(3)symP}
\bigl[\, \vec{\phi}'_1, \; \vec{\phi}'_2, \; \vec{\phi}'_3 \,\bigr](x,y,\tau)=\bigl[\, -\vec{\phi}_1, \; -\vec{\phi}_3, \; -\vec{\phi}_2 \,\bigr](-x,y,\tau) \;.
\end{equation}
\item $2\pi/6$ rotation symmetry (around the $\alpha=1$ vertex),
\begin{equation}  \label{SU(3)symR}
\bigl[\, \vec{\phi}'_1, \; \vec{\phi}'_2, \; \vec{\phi}'_3 \,\bigr](x,y,\tau)=\bigl[\, -\vec{\phi}_1, \; -\vec{\phi}_3, \; -\vec{\phi}_2 \,\bigr](x',y',\tau) \;,
\end{equation}
with  $x'=(x-\sqrt{3}y)/2,\; y'=(\sqrt{3}x+y)/{2}$.
\item Time reversal symmetry can be most easily seen in the real-time Lagrangian $\mathscr{L}_\mathrm{R}$, and it is generated by the flip of the real time $t\mapsto -t$ with the complex conjugation of the coupling constants. 
\end{enumerate}
In the above, we neglect the higher order correction in the lattice spacing. 

Let us make several remarks. 
We here include the gauge redundancy in the list  of symmetry as it is convenient to constrain the possible Lagrangian, although it is not a genuine symmetry of the system and does not act on physical operators. 
Gauge symmetry is obvious in the expression of eq.~\eqref{SU(3)effactionA}, where gauge fields $a_{\alpha, \mu}$ transform as $a_{\alpha, \mu}'=a_{\alpha, \mu}-\partial_\mu\vartheta_\alpha$. 

On fields $\vec{\phi}_{\alpha}$, the spin rotation acts as if it is an $SU(3)$ transformation, $\vec{\phi}_{\alpha}\to V\vec{\phi}_{\alpha}$.  
However, the transformation by the center element, $V=\omega I_3$ ($\omega=\exp(2\pi \im/3)$), is a part of the $U(1)\times U(1)$ gauge transformation, so the center subgroup should be eliminated from the global symmetry. 
In other words, for any gauge-invariant operators, the spin rotation symmetry acts as $PSU(3)=SU(3)/\mathbb{Z}_3$. 

$\mathbb{Z}_3$ cyclic symmetry corresponds to the invariance of the translation by one lattice spacing.
When we perform the lattice translation by one unit, then the red, green, and blue sublattices of Fig.~\ref{fig:SU(3)coordinate} are cyclically permuted. 
Since the label $\alpha$ of $\vec{\phi}_{\alpha}$ is nothing but the assignment of spin vectors to a sublattice, the label is cyclically permuted by the lattice translation. 
Combined with the $x$-axis reflection and the $2\pi/6$ rotation, we can generate the whole $p6m$ wallpaper group, so these are the sufficient condition to achieve the underlying lattice symmetry.

We have derived the sigma model Lagrangian~\eqref{SU(3)effaction} starting from  the lattice model by using the semiclassical approximation in Sec.~\ref{sec:NLS}. 
It turns out that the Lagrangian~\eqref{SU(3)effaction} is the unique possibility that can be constructed by $\vec{\phi}_{\alpha}$ and satisfies the whole symmetry within the quadratic orders of the spacetime derivatives. 
For the proof, see Appendix~\ref{sec:B}. 

In the following of this section, let us discuss the emergent symmetry of the $SU(3)/U(1)^2$ sigma model. 
We first note that the spatial rotation is enlarged to the full $SO(2)$ symmetry, and moreover, the theory acquires the relativistic $SO(3)$ invariance in the Euclidean signature. 
As a result, the $2\pi/6$ rotation~\eqref{SU(3)symR} can be regarded as an internal symmetry, $\vec{\phi}_2\leftrightarrow \vec{\phi}_3$, up to an overall sign. 
Combined with the $(\mathbb{Z}_3)_{\mathrm{trans}}$ symmetry, the effective theory has the symmetric group $S_3=(\mathbb{Z}_3)_{\mathrm{trans}}\rtimes (\mathbb{Z}_2)_{\mathrm{rot}}$ as an internal symmetry, 
\begin{equation}
    \vec{\phi}_\alpha\mapsto \mathrm{sign}(s)\vec{\phi}_{s(\alpha)} \;,
\end{equation}
with $s \in S_3$ and $\mathrm{sign}(s)=\pm 1$ depending on whether $s$ is an even or odd permutation. 
The negative sign for the odd permutation is required to keep the condition $\vec{\phi}_1\cdot (\vec{\phi}_2\times \vec{\phi}_3)=1$ invariant under the transformation. 
Although this $S_3$ permutation has a well-established origin as the lattice symmetry, we need the emergent continuous $SO(2)(\subset SO(3))$ spacetime rotational invariance to regard this as an internal symmetry. 

In addition, there is also a topological $U(1)\times U(1)$ symmetry in the $3$d $SU(3)/U(1)^2$ sigma model. 
The Noether current $j_{\alpha}=j_{\alpha,\mu}\diff x^\mu$ is given by 
\begin{equation}  \label{SU(3)symTopcurrent}
j_{\alpha}=\star {1\over 2\pi}\diff a_{\alpha} \;, 
\end{equation}
where $\star$ denotes the Hodge dual operation. 
The conservation law, $\partial^\mu j_{\alpha,\mu}=0$, comes out of the Bianchi identity, 
\begin{equation}
    \diff \star j_{\alpha}=\diff \left({1\over 2\pi}\diff a_{\alpha}\right)=0 \;,  
\end{equation}
and it generates the $U(1)$ symmetry, whose periodicity is given by the Dirac quantization condition, 
\begin{equation}
    \int_{M_2} \star j_{\alpha}={1\over 2\pi}\int_{M_2}\diff a_{\alpha}\in \mathbb{Z} \;, 
\end{equation}
for any closed surfaces, $M_2$. 
When we take $M_2$ as the spatial manifold, they are nothing but the charges of magnetic skyrmions. 
Since 
\begin{equation}
    \sum_{\alpha} j_{\alpha}=0 \;, 
\end{equation}
only two of them are independent, which means that the system has two different $U(1)$ charges for skyrmions. 
This is consistent with the homotopy computation, $\pi_2(SU(3)/U(1)^2)\simeq \mathbb{Z}^2$, which also suggests the existence of two conserved charges. 

However, the lattice model~\eqref{hamiltonian} does not have the corresponding $U(1)$ symmetries, which means that this topological $U(1)\times U(1)$ symmetry is an emergent symmetry of the sigma-model description. 
There should be some high-energy processes at the lattice scale which explicitly breaks this symmetry, so that skyrmion numbers jump. 
In order to take this effect into account, we must intoduce the defect operator, called the monopole, which is defined by a singularity, at which 
\begin{equation}
    \int_{S^2_*}\diff a_{\alpha}\not=0 \;,
\end{equation}
where $S^2_*$ is the small sphere surrounding the singular point. 
The liberation of dynamical monopoles leads to the confinement of compact $U(1)$ gauge theory~\cite{Polyakov:1976fu, Polyakov:1987ez}, which can have a different behavior from that of the N\'eel phase corresponding to the Higgs regime. 
It is quite important to understand the semiclassical properties of dynamical monopoles in order to discuss the phase transition of \eqref{hamiltonian} using QFT, and we consider such processes in the next section.

\section{Berry phase of monopoles}  \label{sec:tunnel}

In this section, we consider the tunneling process that changes the skyrmion numbers, which are called the monopole event. 
In the previous section~\ref{sec:symmetry}, we have seen that the $SU(3)/U(1)^2$ sigma model enjoys the emergent symmetry, $U(1)^2_{\mathrm{top}}\rtimes S_3$, but some high-energy processes should break a part of it explicitly. 
By computing the monopole-related amplitudes, we shall identify how the exact symmetry is embedded into the emergent one, which gives the constraint on the possible perturbations of the effective Lagrangian.

\subsection{Skyrmion configurations}  \label{ssec:hedgehog}

Since monopoles are defined by the process that changes the skyrmion numbers, we should construct the explicit form of skyrmion configurations before discussing monopoles. 
For this purpose, let us first introduce a local coordinate for the spin coherent vectors $\vec{\Phi}_{\alpha}$ of sublattices $\alpha=1,2,3$ as follows:
\begin{align}
\vec{\Phi}_1 &= \left( \begin{array}{c}
\sin\theta \cos\varphi \\
\eue{\im \alpha_2}\sin\theta\sin\varphi \\
\eue{\im \alpha_3}\cos\theta
\end{array} \right) \;,  \label{SU(3)para1} \\
\vec{\Phi}_2 &= \cos\gamma \left( \begin{array}{c}
\cos\theta\cos\varphi \\
\eue{\im\alpha_2}\cos\theta\sin\varphi \\
-\eue{\im\alpha_3}\sin\theta
\end{array} \right) + \eue{\im\delta}\sin\gamma \left( \begin{array}{c}
\sin\varphi \\
-\eue{\im\alpha_2}\cos\varphi \\
0
\end{array} \right) \;,  \label{SU(3)para2} \\
\vec{\Phi}_3 &= \sin\gamma \left( \begin{array}{c}
\cos\theta\cos\varphi \\
\eue{\im\alpha_2}\cos\theta\sin\varphi \\
-\eue{\im\alpha_3}\sin\theta
\end{array} \right) - \eue{\im\delta}\cos\gamma \left( \begin{array}{c}
\sin\varphi \\
-\eue{\im\alpha_2}\cos\varphi \\
0
\end{array} \right),  \label{SU(3)para3}
\end{align}
with the following six parameters,
\begin{equation}  \label{parameterdomain}
\theta\,,\,\varphi\,,\,\gamma \in \left[0,\,\tfrac{\pi}{2}\right] \;\;\mathrm{and}\;\; \alpha_2\,,\,\alpha_3\,,\,\delta \in \left[0,\,2\pi\right) \;. 
\end{equation}
Let us explain our physical motivation of this specific choice of parameters. 
So far, we have seen that the $SU(3)/U(1)^2$ sigma model is a good description of the N\'eel phase. 
When we assume that the above parameters, $(\theta,\varphi,\gamma,\alpha_2,\alpha_3,\delta)$, depends only on the position of the unit cell $(j,k)$ but not on the sublattice label $\alpha$, 
then $(\vec{\Phi}_1,\vec{\Phi}_2,\vec{\Phi}_3)\in SU(3)$, so we can regard them as a local coordinate of the sigma-model target space $SU(3)/U(1)^2$.  
Indeed, this explicit form can be obtained basically by fixing two $U(1)$ phases in the local coordinate of the $SU(3)$ matrix given in Ref.~\cite{Bronzan:1988wa} with some redefinition of parameters. 

We can relax the condition for $SU(3)/U(1)^2$ by allowing that the parameters $(\theta, \varphi, \gamma, \alpha_2, \alpha_3, \delta)$ depends also on the sublattice label $\alpha$. 
As long as they smoothly vary with $(j,k,\alpha,\tau)$, Eqs.~\eqref{SU(3)para1}--\eqref{SU(3)para3} can describe the small fluctuation from the N\'eel order. In this case, three coherent vectors in a unit cell are no longer orthogonal, and we just find that $\vec{\Phi}_\alpha(j,k,\tau) \in \mathbb{C}P^2$. Actually, we can regard that the six parameters above are the $SU(3)$ generalization of the staggered magnetic order parameter $\vec{m}=(-1)^{x+y}\vec{n}=(\sin\theta\cos\varphi \,,\, \sin\theta\sin\varphi \,,\, \cos\theta) \in S^2$ for the $SU(2)$ antiferromagnets, when $\vec{S}=s\,\vec{n}$ describes the original spins and $(x,y)\in \mathbb{Z}^2$ labels the square lattice.

Using the coordinates~\eqref{SU(3)para1}--\eqref{SU(3)para3}, we construct magnetic skyrmions, which have non-trivial topological charges $Q_\alpha^{xy}$. 
Although our construction does not exhaust all the skyrmion configurations, we obtain sufficiently many examples to obtain the symmetry property of monopoles. 
By the analogy of the hedgehog configuration in $SU(2)$ spins~\cite{Haldane:1988zz, Read:1989jy}, we regard $(2\theta, 2\varphi, 2\gamma)$ as polar angles, and $(\alpha_2, \alpha_3, \delta)$ as azimuthal angles. 
Introducing the polar coordinate $(r,\eta)$ of the $xy$ plane, $x+\im y=r\rme^{\im \eta}$, a hedgehog configuration centered at origin can be obtained by the following procedures:
\begin{itemize}
\item some of $(\theta, \varphi, \gamma)$ are set to $f(r)$ or $\tfrac{\pi}{2}-f(r)$ , and
\item $(\alpha_2,\alpha_3,\delta)=(\ell\, \eta, m\, \eta, n\, \eta)$ up to constant shifts with $\ell,m,n\in \mathbb{Z}$ \;,
\end{itemize}
where $f(r)$ is an arbitrary function which connects $f(0)=\tfrac{\pi}{2}$ and $f(\infty)=0$ smoothly. 
As an example, one can choose 
$f(r)=\frac{1}{2}\arccos\left(\frac{2(r/r_0)^2-1}{2(r/r_0)^2+1}\right)$, where $r_0$ controls the size of skyrmions. 
In order for this configuration to be well-defined in the continuum limit, we have to impose the following condition:
\begin{itemize}
\item Any components in $\vec{\Phi}_\alpha$ of the form $R(f(r))\rme^{\im\eta}$ must satisfy $|R(f(0))|=0$ (after suitable $U(1)$ gauge transformations). The same should be true also for $r=\infty$ so that we can compactify the spatial manifold. 
\end{itemize}
In Table~\ref{tab:skyrmion}, we give a list for examples of the skyrmion configuration. 
Let us pick the uppermost of the table for the explanation: we set $\theta=f(r)$, $\alpha_2=m\eta$, and others to be $0$, and we get 
\begin{equation}
    \vec{\Phi}_1=\begin{pmatrix}
    \sin f(r)\\
    0\\
    \rme^{\im m \eta}\cos f(r)
    \end{pmatrix}, \,\,
    \vec{\Phi}_2=\begin{pmatrix}
    \cos f(r)\\
    0\\
    -\rme^{\im m \eta}\sin f(r)
    \end{pmatrix},     \,\,
    \vec{\Phi}_3=\begin{pmatrix}
    0\\
    1\\
    0
    \end{pmatrix} \;. 
\end{equation}
Let us check that this satisfies the required condition. Since $f(0)=\pi/2$, $\vec{\Phi}_1$ is well-defined at $r=0$, while $\vec{\Phi}_2$ looks to be ambiguous at the coefficient of $\rme^{-\im m \eta}$ remains finite. 
However, we can use the $U(1)$ gauge transformation, so that 
\begin{equation}
    \vec{\Phi}'_2=\rme^{-\im m \eta}\vec{\Phi}_2=\begin{pmatrix}
    \rme^{-\im m \eta}\cos f(r)\\
    0\\
    -\sin f(r)
    \end{pmatrix}\;, 
\end{equation}
which is well-defined at $r=0$. 
For $r=\infty$, $\vec{\Phi}_2$ is well-defined, and $\vec{\Phi}_1$ is also well-defined after the gauge transformation, so that we can regard this as a configuration on $S^2$. 
This configuration is an embedding of the $SU(2)/U(1)$ magnetic skyrmion of the usual $SU(2)$ antiferromagnets into the $SU(3)/U(1)^2$ sigma model. 
\begin{table} 
  \centering
  \begin{tabular}{ccccccc}
	\hline
	$\theta$ & $\varphi$ & $\gamma$ & $\alpha_2$ & $\alpha_3$ &  $\delta$ & $(Q_1^{xy},\,Q_2^{xy},\,Q_3^{xy})$ \\
	\hline \hline
	$f(r)$ & 0 & 0 & const. & $m\eta$ &  const. & $(-m,m,0)$ \\
	$f(r)$ & $\pi$/2 & 0 & $\ell \eta$ & const. & const. & $(\ell,-\ell,0)$ \\
	$\pi$/2 & $f(r)$ & 0 & $\ell\eta$ & const. & const. & $(\ell,0,-\ell)$ \\
	0 & 0 & $f(r)$ & const. & const. & $n\eta$ & $(0,n,-n)$ \\
	$f(r)$ & $f(r)$ & 0 & $\ell \eta$ & $m\eta$ & const. & $(\ell-m,m,-\ell)$ \\
	$f(r)$ & $\pi$/2 & $f(r)$ & $\ell\eta$ & $m\eta$ & $n\eta$ & $(\ell-m,n-\ell,m-n)$ \\
	$f(r)$ & $f(r)$ & $f(r)$ & $\ell\eta$ & $m\eta$ & $n\eta$ & $(\ell-m,n,-\ell+m-n)$ \\
	\hline
  \end{tabular}  
   \caption{Examples of skyrmion configuration. Here, $(\ell,m,n)$ are the set of integers, which characterizes the topological charges. }
  \label{tab:skyrmion}
\end{table}

Let us calculate topological charges of skyrmions. 
Using Eqs.~\eqref{SU(3)para1}--\eqref{SU(3)para3}, we first compute the corresponding $U(1)$ gauge fields, $\calA_{\alpha}=\im \vec{\Phi}^*_\alpha \cdot \diff \vec{\Phi}_\alpha$, and we obtain 
\begin{align}
\left(\begin{array}{c}
\calA_{1}\\
\calA_{2}\\
\calA_{3}
\end{array}\right)
&= -\left[ \diff \delta
\left( \begin{array}{c} 0 \\ + \\ - \end{array} \right)
\sin^2\gamma + \diff {\varphi}
\left( \begin{array}{c} 0 \\ + \\ - \end{array} \right) 
2\sin\delta\sin\gamma\cos\gamma\cos\theta \right. \nonumber \\
&\quad + \diff{\alpha}_2\left\{
\left( \begin{array}{c} 0 \\ - \\ + \end{array} \right) 
\sin^2\gamma\cos^2\theta\sin^2\varphi + 
\left( \begin{array}{c} + \\ - \\ 0 \end{array} \right) 
\sin^2\theta\sin^2\varphi \right. \nonumber \\
&\qquad\qquad\left. + 
\left( \begin{array}{c} 0 \\ - \\ + \end{array} \right) 
\cos^2\gamma\cos^2\varphi +
\left( \begin{array}{c} 0 \\ - \\ + \end{array} \right) 
2\cos\delta\sin\gamma\cos\gamma\cos\theta\sin\varphi\cos\varphi \right\} \nonumber \\
&\quad\left. + \diff{\alpha}_3 \left\{ 
\left( \begin{array}{c} 0 \\ - \\ + \end{array} \right) 
\sin^2\gamma\sin^2\theta + 
\left( \begin{array}{c} - \\ + \\ 0 \end{array} \right) 
\sin^2\theta \right\} \right] 
\:,  \label{Berrytunnel}
\end{align}
up to $U(1)$ gauge transformations. We then find the skyrmion charges by 
\begin{align}
Q_\alpha^{xy} &= \frac{1}{2\pi}\int_{\mathbb{R}^2_{xy}}\diff \calA_{\alpha} \;.  \label{Skycharge3}
\end{align}
Since $Q_\alpha^{xy}\in\mathbb{Z}$ in the continuum limit, any smooth deformations of the configuration do not change them. 
This can be also understood from the fact that we can identify $\calA_{\alpha}$ with $a_{\alpha}=\im \vec{\phi}^*_\alpha\cdot \diff \vec{\phi}_\alpha$ as long as the continuum approximation is valid. 
In Table~\ref{tab:skyrmion}, we also give the list of the skyrmion charges for each example. 

The mass of skyrmion, $\mathcal{M}_{\mathrm{skyrmion}}$, can also be evaluated by substituting $\vec{\phi}_\alpha=\vec{\Phi}_\alpha$ into the effective Lagrangian~\eqref{SU(3)effaction}:
\begin{equation}
    \mathcal{M}_{\mathrm{skyrmion}}=\sum_{\alpha=1}^{3}\int\diff x \diff y \,{1\over g_{\mathrm{eff}}}\sum_{\mu=x,y}|(\partial_\mu+\im \calA_{\alpha,\mu})\Phi_\alpha|^2. 
\end{equation}
There is the Bogomol'nyi-Prasad-Sommerfield (BPS) bound~\cite{Bogomolny:1975de,Prasad:1975kr} for the mass of skyrmions (see  Ref.~\cite{Hongo:2018rpy}):
\begin{equation}
    \mathcal{M}_{\mathrm{skyrmion}}\ge {2\pi\over g_{\mathrm{eff}}}\sum_{\alpha=1}^3 |Q_\alpha^{xy}| \;. 
\end{equation}
The equality holds if and only if the (anti-)BPS equations for $\alpha=1,2,3$ can be satisfied simultaneously while keeping the orthonormality constraint. 
Especially when we embed the $\mathbb{C}P^1$ skyrmion into $SU(3)/U(1)^2$, we can see that the (anti-)BPS equation is satisfied with keeping the orthonormality, and thus the BPS bound is saturated. 
For such solutions, there is a size moduli, i.e., skyrmions with different size have the same mass as in the case of $\mathbb{C}P^1$ skyrmions. 
For other types of skyrmions, where all three topological charges are nonzero, satisfying the BPS bound seems to be difficult.

\subsection{Wess-Zumino term and Berry phase of monopoles}  \label{ssec:berrym}

Although the skyrmion numbers are conserved quantities in the continuum approximation, there are tunneling processes at the lattice scale between states with different topological charges. 
In the context of low-energy field theory, such a process is described by insertions of defect operators, which are called monopoles in the description of $U(1)$ gauge theories. 
To consider the skyrmion-number changing processes, we have to introduce the perturbation of the effective Lagrangian by those monopole operators. 
For this purpose, we would like to know what kind of monopole operators can be added to the effective Lagrangian, and this requires us to identify how the symmetry acts on those monopoles. 

Following the seminal work by  Haldane~\cite{Haldane:1988zz}, we compute the Berry phase of monopoles by constructing an explicit lattice process changing the skyrmion numbers. 
As we have seen in Sec.~\ref{sec:NLS}, the complex phase in the path integral completely disappears in the continuum approximation, even though the Wess-Zumino term could give an imaginary part to the Euclidean action. 
This is true not only for small perturbations around the classical N\'eel state but also for any states with nonzero skyrmion charges, so long as the continuum approximation is valid for any spacetime points. 
When monopoles are present, however, there are spacetime points, at which the continuum approximation breaks down, and there can be a nonzero imaginary contribution from the Wess-Zumino term at those points~\cite{Haldane:1988zz, Read:1989zz, Read:1989jy, Read:1990zza}. 
This gives important information to constrain the possible monopole perturbations.

We here use the skyrmion listed in the second from the bottom of Tabble~\ref{tab:skyrmion} as an example. 
Correspondingly, we construct an explicit lattice process, where the topological charges change as
\begin{equation}  \label{tunnelingexample}
\begin{pmatrix}
Q_1^{xy}(\tau)\\ Q_2^{xy}(\tau) \\ Q_3^{xy}(\tau)
\end{pmatrix}:
\begin{pmatrix}
0\\ 0\\ 0
\end{pmatrix} \;\to\; 
\begin{pmatrix}
Q_1\\
Q_2\\
Q_3
\end{pmatrix}
\equiv 
\begin{pmatrix}
\ell-m\\
n-\ell\\
m-n
\end{pmatrix}
\;\to\; 
\begin{pmatrix}
0\\ 0\\ 0
\end{pmatrix} \;, 
\end{equation}
for $\tau:0\to \beta$,  
and calculate its Berry phase. 
$Q_1,Q_2,Q_3$ denote the topological sector of the skyrmion, which is created and annihilated at $\tau=T_1$ and $T_2$, respectively, and only two of them are independent, $Q_1+Q_2+Q_3=0$. 
As we have done in Sec.~\ref{ssec:hedgehog}, we put the following ansatz,  
\begin{equation}  \label{Skylm}
\theta=\gamma=F(\vec{r},\tau) \:,\; \varphi=\frac{\pi}{2} \:,\; 
(\alpha_2, \alpha_3, \delta)=(\ell \,\eta(\vec{r},\tau),m \,\eta(\vec{r},\tau),n \,\eta(\vec{r},\tau)) \;,  
\end{equation}
where $F$ and $\eta$ are functions of $(\vec{r},\tau)=(x,y,\tau)$. 
The spin coherent vectors~\eqref{SU(3)para1}--\eqref{SU(3)para3} become 
\begin{equation}
    \vec{\Phi}_1=\begin{pmatrix}
    0\\
    \rme^{\im \ell \eta}\sin F\\
    \rme^{\im m \eta} \cos F
    \end{pmatrix},\quad 
    \vec{\Phi}_2=\begin{pmatrix}
    \rme^{\im n \eta} \sin F\\
    \rme^{\im \ell \eta}\cos^2 F\\
    -\rme^{\im m \eta} \sin F\cos F
    \end{pmatrix}, \quad 
    \vec{\Phi}_3=\begin{pmatrix}
    -\rme^{\im n \eta}\cos F\\
    \rme^{\im \ell \eta}\sin F\cos F\\
    -\rme^{\im m \eta} \sin^2 F
    \end{pmatrix},
\end{equation}
and the corresponding $U(1)$ gauge fields~\eqref{Berrytunnel} are 
\begin{equation}
    \begin{pmatrix}
    \calA_1\\
    \calA_2\\
    \calA_3
    \end{pmatrix}
    =-\diff \eta \left[
    \sin^2 F\begin{pmatrix}
    \ell-m\\
    n+m-2\ell\\
    \ell-n
    \end{pmatrix}
    +\sin^4 F \begin{pmatrix}
    0\\
    \ell-m\\
    m-\ell
    \end{pmatrix}
    \right]. 
\end{equation}
By mapping $(x,y)\to (j,k,\alpha)$, we calculate the Berry phase of this process as follows:
\begin{align}
    S_{\mathrm{B}}&=\im\, p \sum_{(j,k,\alpha)}\Omega(j,k,\alpha) \;, \nonumber\\
    \Omega(j,k,\alpha)&=\int_{0}^\beta (-\calA_{\alpha,\tau}(j,k)) \,\diff \tau \;. 
    \label{BerryOmega}
\end{align} 
To obtain $\mathcal{A}_\alpha$, we have performed the $U(1)$ gauge transformation, so that it is proportional to $\sin^2 F$. 
It changes the Berry phase only by $2\pi \im\, p\,\mathbb{Z}$, which does not affect the path integral. 
By requiring that $F(\vec{r},\tau) \to 0$ as $|\vec{r}|\to \infty$, the $U(1)$ gauge fields disappear at spatial infinities, $\calA_\alpha \to 0$, and thus the Berry phase does not receive any contributions from the spatial infinities, so we do not need to care about the boundary condition for the computation of $S_{\mathrm{B}}$.

\begin{figure} \centering
\includegraphics[scale=0.32]{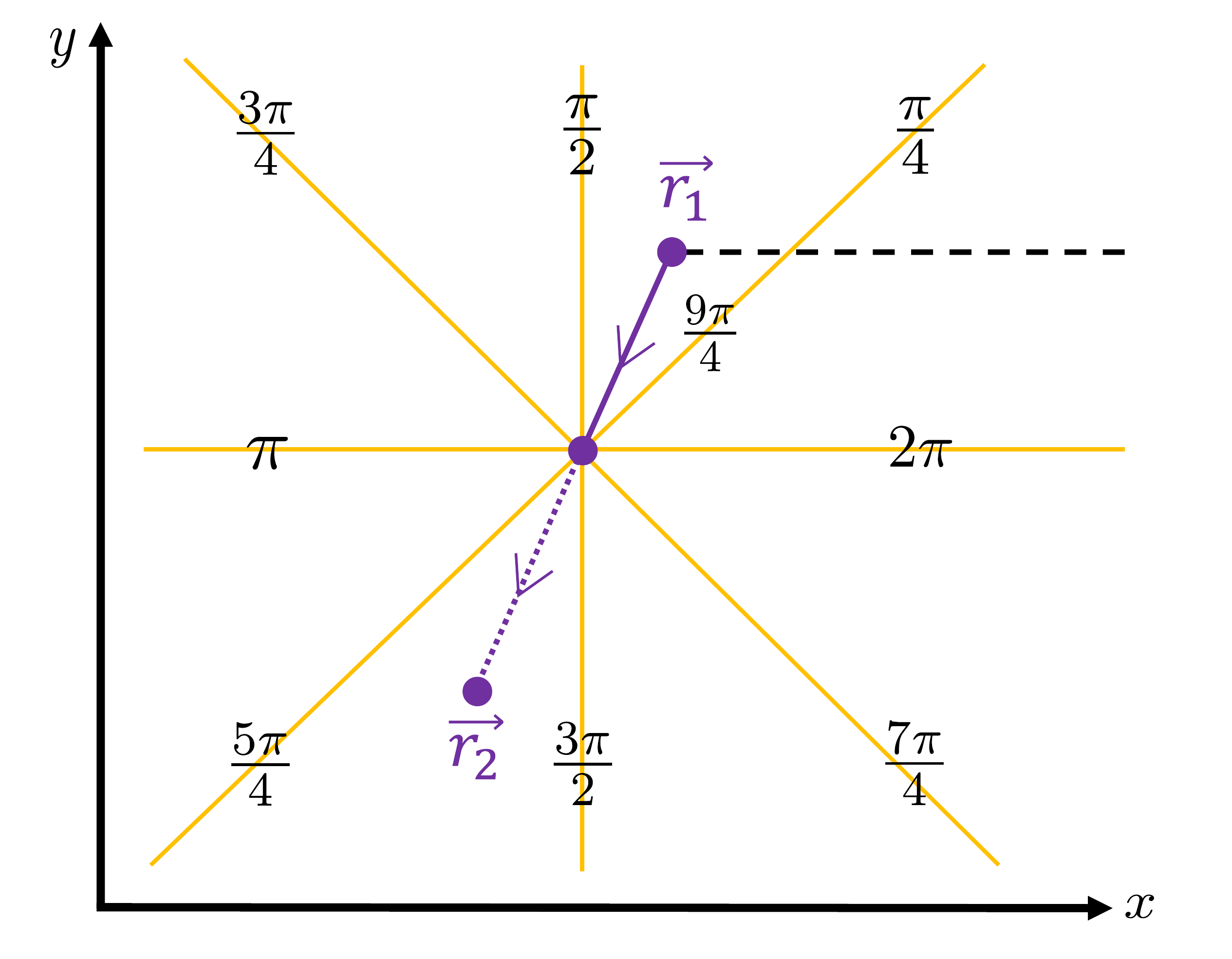}
\caption{Definition of principal value of the argument $\eta_{12}(\vec{r},\tau)$. 
Here, the skyrmion is created at the location $\vec{r}_1$ at the time $T_1$,  moves from $\vec{r}_1$ to $\vec{r}_2$, and is annihilated at $\vec{r}_2$ at the time $T_2$, and we here pick a time $\tau$ between $T_1<\tau<T_2$. 
To make the change of $\eta_{12}(\vec{r},\tau)$ continuously, we first fix the branch cut (shown with dashed line) when the skyrmion is created, and impose that $\eta_{12}$ always jumps by $2\pi$ on the branch cut. 
As $\eta_{12}(\vec{r},\tau)$ denotes the argument of $\vec{r}$ measured from the location of the skyrmion at the time $\tau$, 
there have to be additional $2\pi$ jumps in $\eta_{12}$ along the path of the singular point. 
%when the skyrmion moves
That is the discontinuity line, which is illustrated with the arrowed purple line connecting $\vec{r}_1$ and $\vec{r}_2$. }
\label{fig:principalArg}
\end{figure}

The tunneling process~\eqref{tunnelingexample}, in which a skyrmion appears at $(x_1,\,y_1,\,T_1)$ and disappears at $(x_2,\,y_2,\,T_2)$, is expressed by the following four steps (i)--(iv):
\begin{align}
&\text{(i)}\, 0\leq\tau<T_1:
&&\left\{\begin{array}{l}
F(\vec{r},\tau)=\dfrac{\tau}{T_1}f(r_1(\vec{r})) \;,\\ 
\eta(\vec{r},\tau)=\eta_1(\vec{r}) \;,
\end{array}
\right. \nonumber\\
&\text{(ii)}\,  T_1\leq\tau<T_2:
&&\left\{\begin{array}{l}
F(\vec{r},\tau)=f\bigl(r_{12}(\vec{r},\tau)\bigr) \;,\\
\eta(\vec{r},\tau)=\eta_{12}(\vec{r},\tau) \;,
\end{array}\right.
\nonumber\\
&\text{(iii)}\, T_2\leq\tau<T_3:
&&\left\{\begin{array}{l}
F(\vec{r},\tau)=\dfrac{T_3-\tau}{T_3-T_2}f(r_2(\vec{r})) \;, \\
\eta(\vec{r},\tau)=\eta_2(\vec{r}) \;,
\end{array}\right.
\nonumber\\
&\text{(iv)}\, T_3\leq\tau\leq\beta:
&&\left\{\begin{array}{l}
F(\vec{r},\tau)=0 \;, \\ 
\eta(\vec{r},\tau)=\eta_2(\vec{r})+\dfrac{\tau-T_3}{\beta-T_3}(\eta_1(\vec{r})-\eta_2(\vec{r})) \;,
\end{array}\right.
\end{align}
where $f(r)$ is a smooth function with $f(0)=\pi/2$, $f(\infty)=0$, and  we have defined 
\begin{align}
    r_1\rme^{\im \eta_1}=(x-x_1)+\im (y-y_1) \;,\quad 
    r_2\rme^{\im \eta_2}=(x-x_2)+\im (y-y_2) \;,
\end{align}
and 
\begin{align}
    r_{12}\rme^{\im \eta_{12}}=
    \left(x-x_1-\frac{\tau-T_1}{T_2-T_1}(x_2-x_1)\right)
    +\im \left(y-y_1-\frac{\tau-T_1}{T_2-T_1}(y_2-y_1)\right) \;. 
\end{align}
The principal value of the argument $\eta_{12}$ is defined so that it varies continuously during the process, as shown in Fig.~\ref{fig:principalArg}, and this introduces the \textit{discontinuity line}~\cite{Haldane:1988zz} shown by the arrowed purple line. 
In step~(i) and (iii), the spin coherent vectors $\vec{\Phi}_\alpha$ have the conical singularity at $\vec{r}_1$ and $\vec{r}_2$, respectively, so we cannot take the naive continuum limit of this process. 
Step~(iv) is necessary to achieve the periodic boundary condition.\footnote{
In our specific choice here, the Berry phase do not get any contributions from step~(iv) as $\calA_{\alpha}=0$ for $F=0$. 
However, this is somewhat accidental. 
For example, the tunneling process between $(\theta,\varphi,\gamma,\alpha_2,\alpha_3,\delta)=(\tfrac{\pi}{2},0,0.\eta,0,0)$ and $(\tfrac{\pi}{2},f,0,\eta,0,0)$, which corresponds to the third from the uppermost of Table~\ref{tab:skyrmion}, has a finite contribution from step~(iv). } 

\begin{figure} \centering
\includegraphics[scale=0.32]{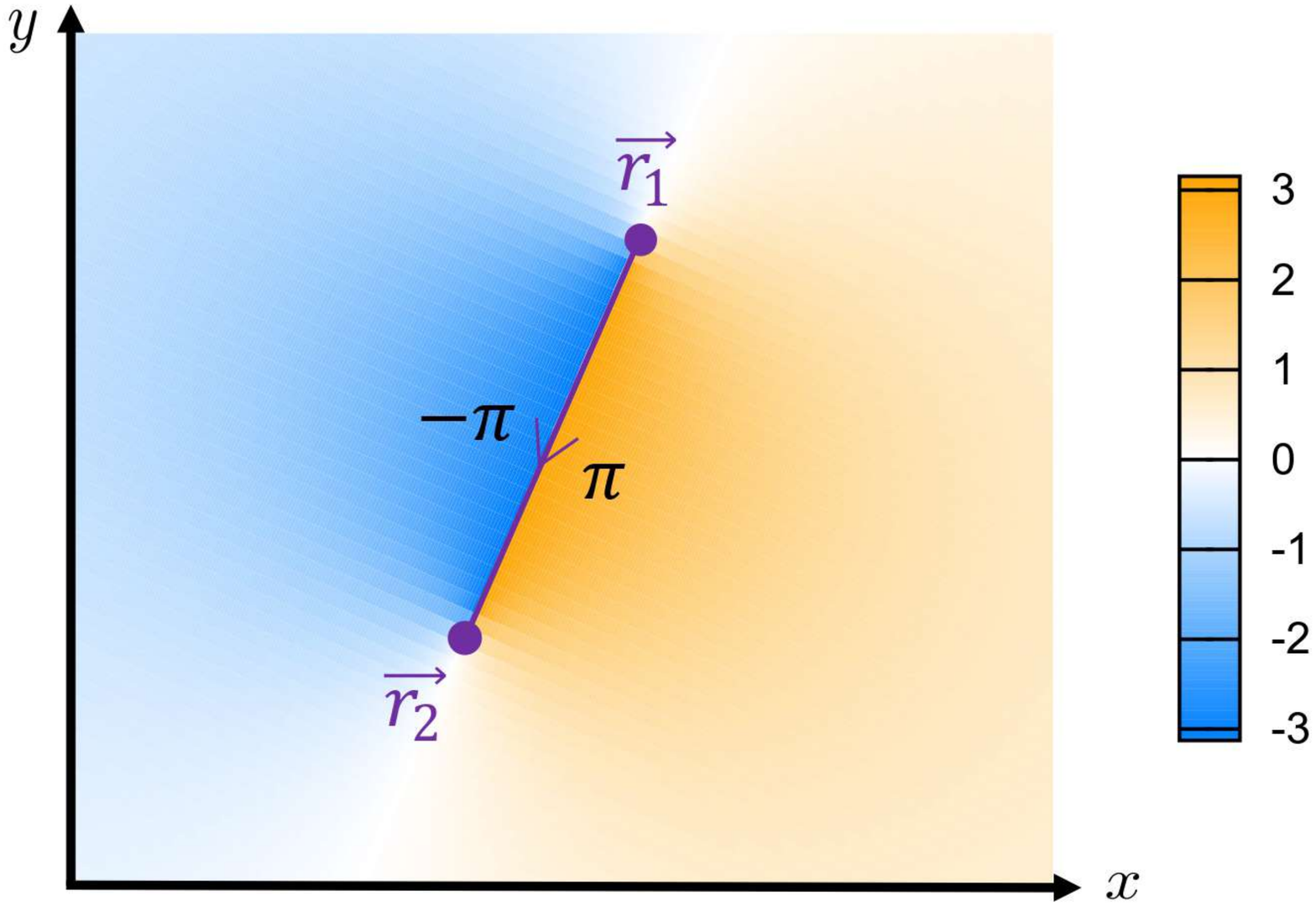}
\caption{The value of $\left[\mathrm{arctan}\frac{y-y_2}{x-x_2}-\mathrm{arctan}\frac{y-y_1}{x-x_1}\right]$ in our convention. The discontinuity line is illustrated as purple bold line between $\vec{r}_1$ and $\vec{r}_2$, which corresponds to the path of the skyrmion. The movement of the skyrmion in $\tau$-direction is projected in $xy$-plane. }
\label{fig:discontinuity}
\end{figure}

Since we need $\dot{\eta}=\diff \eta/\diff \tau \not= 0$ for nonzero contributions to the Berry phase~\eqref{BerryOmega}, the step~(i) and (iii) do not contribute, and only the step~(ii) can have a finite contribution:
\begin{equation}
    \Omega(j,k,\alpha) = \int_{T_1}^{T_2}\diff \tau\, \dot{\eta}_{12}\left[
    \sin^2 f(r_{12})\begin{pmatrix}
    \ell-m\\
    n+m-2\ell\\
    \ell-n
    \end{pmatrix}
    +\sin^4 f(r_{12}) \begin{pmatrix}
    0\\
    \ell-m\\
    m-\ell
    \end{pmatrix}
    \right] \;. 
\end{equation}
Since we shall only need the spatial singular behavior that appears when $r_{12}\simeq 0$, we can simply set $\sin^2 f(r_{12})\simeq \sin^2 f(0)=1$ to obtain it.\footnote{If we choose $f(r)=\frac{1}{2}\arccos\left(\frac{2(r/r_0)^2-1}{2(r/r_0)^2+1}\right)$, we have $\sin^2 f(r)=1/(2(r/r_0)^2+1)$ so that the integration can be analytically evaluated to check this claim. }  
We then find that 
\begin{align} 
\Omega(j,k,\alpha) &= \begin{pmatrix}
\ell-m\\
n-\ell\\
m-n
\end{pmatrix} \int_{T_1}^{T_2}\diff \tau \, \dot{\eta}_{12}(\vec{r},\tau)+(\text{reg}.) \nonumber\\
&=
\begin{pmatrix}
Q_1\\
Q_2\\
Q_3
\end{pmatrix}
\left[ \mathrm{arctan}\frac{y-y_2}{x-x_2} - \mathrm{arctan}\frac{y-y_1}{x-x_1}  \right]\:+\; (\text{reg.}) \;,  
\label{Berryexample1} 
\end{align}
where we use $(x,y)\leftrightarrow(j,k,\alpha)$ interchangeably, and ``reg.'' refers the regular piece. 
The branch cut of the first term $\arctan((y-y_2)/(x-x_2))$ consists of the discontinuity line and also of the branch cut of $\arctan((y-y_1)/(x-x_1))$, so \eqref{Berryexample1} has the $2\pi$ jump along the discontinuity line. 
We note that the $2\pi$ jump at the branch cut cancel between two ``$\arctan$'' terms (see Fig.~\ref{fig:discontinuity}). 
Different choice of $f(r)$ only changes the regular part. 
We can notice that the singular part in  \eqref{Berryexample1} is proportional to the charges of the skyrmion, $(Q_1,Q_2,Q_3)=(\ell-m, n-\ell, m-n)$. 
By performing the summation in Eq.~\eqref{BerryOmega}, we can obtain the Berry phase $S_{\mathrm{B}}$ of the monopole events. 

\subsection{Graphical rule for the Berry phase of monopoles}
\label{sec:graphical_rule}

In the following, we obtain the Berry phase of the monopole processes by summing up \eqref{Berryexample1} over lattice sites, $(j,k,\alpha)$. 
It turns out that the computation can be reduced to a simple graphical rule. 
Here, we give a heuristic derivation of the graphical rule to obtain the Berry phase conveniently, following the idea of Haldane for $SU(2)$ spins on the square lattice~\cite{Haldane:1988zz}.\footnote{Even though our derivation is not rigorous, we have checked that our graphical rule is consistent with the numerical computation of the Berry phase. 
The consistency of these calculations is quite surprising to the authors, and, honestly, we do not fully understand why this should be the case. 
Contribution of each site to the Berry phase is completely different in these two approaches, but the final result turns out to be identical. 
It could be understood by insensitivity of the Berry phase to the choice of the skyrmion profile $f(r)$, but such insensitivity seems to be ensured only when the change of $f(r)$ is sufficiently small at the lattice scale. 
} 

We put a technical assumption that the discontinuity line starts and ends at the center of triangles and it does not overlap with the lattice site. 
That is, the center of skyrmion does not overlap with the lattice sites, and it is created and annihilated exactly at the middle of the lattice. 
We now rewrite the summation of \eqref{BerryOmega} over lattice sites into the summation over triangles:
\begin{equation}
    S_{\mathrm{B}}={\frac{\mathrm{i}\,p}{6}}\sum_{\triangle} \Bigl(\Omega(\triangle,1)+\Omega(\triangle,2)+\Omega(\triangle,3)\Bigr) \;, 
    \label{eq:Berry_phase_sum}
\end{equation}
where $\triangle$ refers each unit triangle of the triangular lattice, and $\Omega(\triangle,\alpha)=\Omega(j,k,\alpha)$ for $(j,k,\alpha)\in \triangle$. 
Since each $\triangle$ contains the site of $\alpha=1,2,3$ by once, we just need to specify the value of $\alpha$ to determine $\Omega(\triangle,\alpha)$. 
The overall factor $1/6$ in \eqref{eq:Berry_phase_sum} appears because each site is shared by $6$ triangles. 

We then have to compute contribution of each $\triangle$, 
\begin{equation}
    {1\over 6}\sum_{\alpha=1}^3 \Omega(\triangle, \alpha)={1\over 6}\sum_{(x,y,\alpha)\in \triangle}Q_\alpha\left[ \arctan\frac{y-y_2}{x-x_2} - \arctan\frac{y-y_1}{x-x_1}  \right] \;. 
    \label{eq:Berry_triangle}
\end{equation}
When $\triangle$ is sufficiently far from the discontinuity line, the value of $\arctan$ 
differs only by $\mathcal{O}(a)$ inside the $\triangle$. 
%is almost constant. 
Since $Q_1+Q_2+Q_3=0$, there is no $\mathcal{O}(a^0)$ contribution to the Berry phase, and we may neglect it. 
More rigorously, we can confirm that the sum of contribution from consecutive $6\times 3$ triangles is $\mathcal{O}(a^3)$ by expanding $\Omega$ in power of $a$. Therefore, it is clear that the contribution becomes negligible when the continuum approximation is valid.
%This motivates us to take into account the contribution of $\triangle$ only if the discontinuity line goes across it or terminates in its middle. We neglect all other triangles to construct the graphical rule in the following. 

\begin{figure} \centering
\includegraphics[scale=0.25]{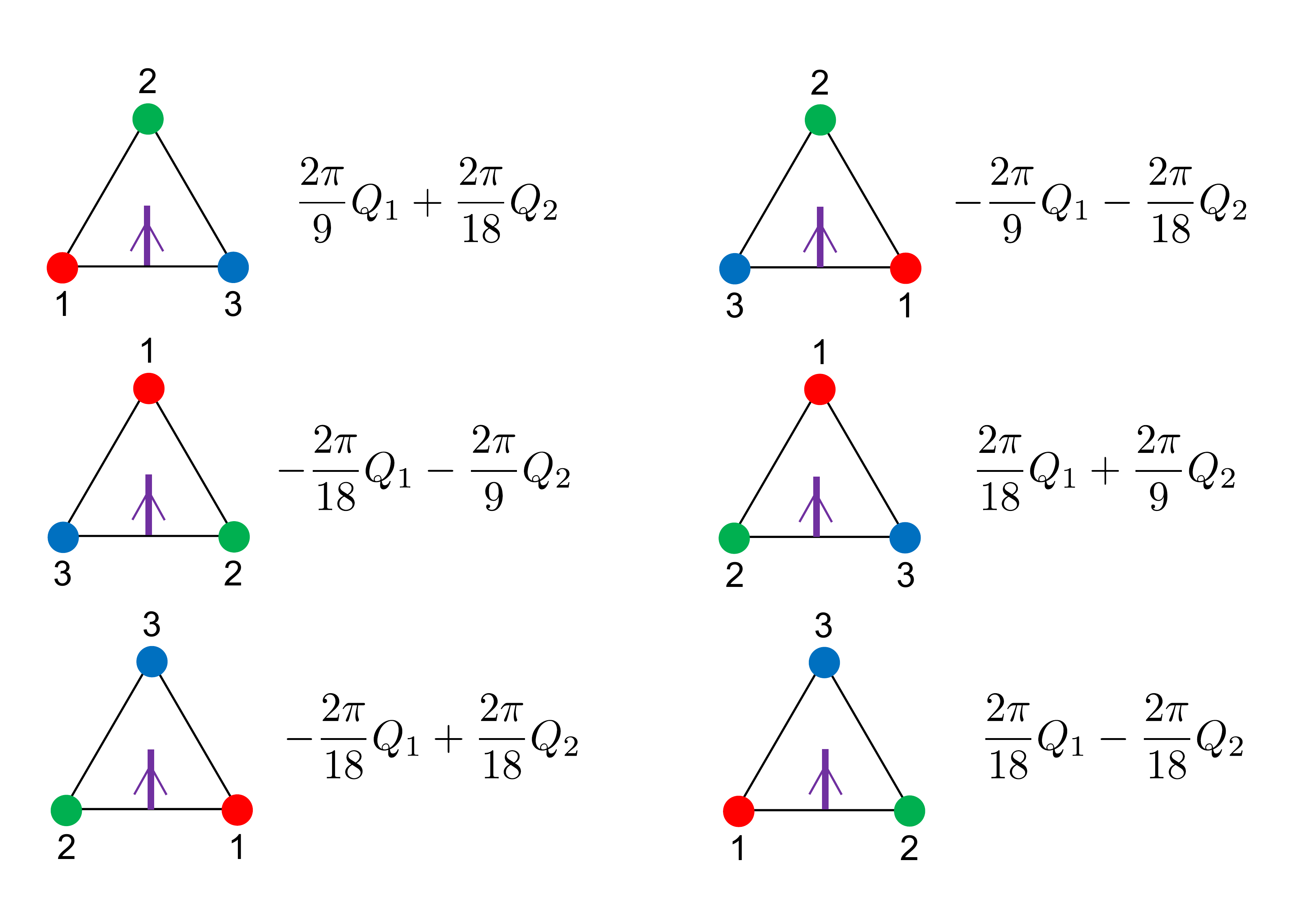}
\caption{Contribution of $\triangle$ to the Berry phase, at which the discontinuity line terminates. We here use $Q_1+Q_2+Q_3=0$ in order to express the results in terms of $Q_1$ and $Q_2$. If the discontinuity line has the opposite orientation, all the signs are flipped. }
\label{fig:newBerry1}
\end{figure}

Let us compute the Berry phase for $\triangle$, at which the skyrmion is annihilated. 
In the actual computation of \eqref{eq:Berry_phase_sum}, the effect of the endpoint of the discontinuity line does not localize to the specific $\triangle$, and it seems that there is no simple rule to find $S_{\mathrm{B}}$ at first sight. 
Still, let us assume that it is possible, and then consider what would be the most appropriate value for \eqref{eq:Berry_triangle}. 
For a concrete computation, we consider the case when the discontinuity line goes across the edge with the $\alpha=1$ (red) and $\alpha=3$ (blue) vertices on the left and the right, respectively (see the upper left figure of Fig.~\ref{fig:newBerry1}). 
Since the $2\pi$ vortex is located at the center of $\triangle$, its contribution to the Berry phase is expressed as
\begin{align}
    {1\over 6}\sum_{\alpha} \Omega(\triangle,\alpha)
    &={1\over 6}\left({2\pi\over 3}Q_1\,+\,0\cdot Q_2+\Bigl(-{2\pi\over 3}\Bigr) Q_3\right)\nonumber\\
    %&={\pi\over 9}(2Q_1+Q_2). 
    &=\frac{2\pi}{9}Q_1+\frac{2\pi}{18}Q_2 \;. 
\end{align}
%If the orientation of the triangle is opposite, we should flip the sign. 
We can compute other triangles in the same way, and we obtain the result shown in Fig.~\ref{fig:newBerry1}. 
When the skyrmion is created on the triangle, we flip their signs.

\begin{figure} \centering
\includegraphics[scale=0.25]{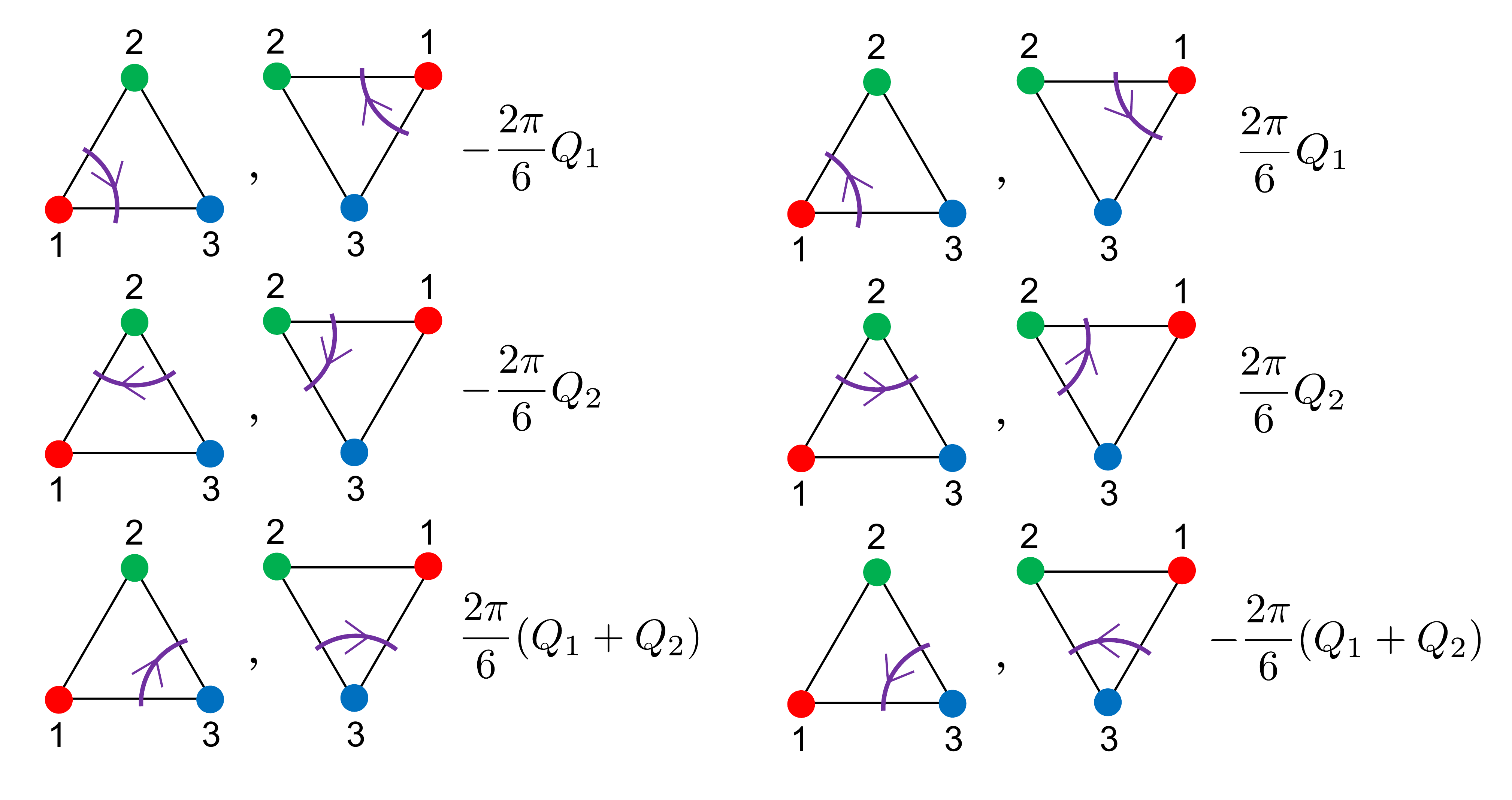}
\caption{Contribution of $\triangle$ to the Berry phase, which is crossed by the discontinuity line. The vertex on the right hand side contributes with $-\pi$, and that of the left hand side contributes with $+\pi$. Using $Q_1+Q_2+Q_3=0$, we write the formula in terms of $Q_1$ and $Q_2$. }
\label{fig:newBerry2}
\end{figure}

Next, we go on to the case where the discontinuity line goes across $\triangle$. 
For that $\triangle$, we have to assign $+\pi$ to the left vertices and $-\pi$ to the right vertices. As an example, let us consider the case when the $\alpha=1$ (red) vertex sits on the left of the discontinuity line and the $\alpha=2,3$ (green, blue) vertices sit on the right (see the upper right figure of Fig.~\ref{fig:newBerry2}), then we find 
\begin{align}
    {1\over 6}\sum_\alpha \Omega(\triangle,\alpha)
    &={\pi\over 6}(Q_1-Q_2-Q_3)\nonumber\\
    &={2\pi\over 6}Q_1 \;. 
\end{align}
The same computations gives the result for other cases shown in Fig.~\ref{fig:newBerry2}. 
Although we have obtained the necessary information to compute the Berry phase, we can obtain more convenient rule by noticing the following equality:
%\begin{align}
%    {2\pi\over 6}Q_1={\pi\over 9}(2Q_1+Q_2)+{\pi\over 9}(Q_1-Q_2). 
%\end{align}
\begin{equation}
\frac{2\pi}{6}Q_1=  \left(\frac{2\pi}{9}Q_1+\frac{2\pi}{18}Q_2\right)+\left(\frac{2\pi}{18}Q_1-\frac{2\pi}{18}Q_2\right) . 
\end{equation}
The right hand side is the sum over the processes computed in Fig.~\ref{fig:newBerry1} with the corresponding crossing of the discontinuity line.  
The first term on the right hand side represents the contribution from the skyrmion that goes across the edge with red-blue vertices and is annihilated at the middle of the triangle. 
The second term represents the contribution from the skyrmion that is created at the middle and goes across the edge with red-green vertices. 
Their sum is equal to the contribution of $\triangle$, which is gone through by the discontinuity line.

\begin{figure} \centering
\includegraphics[scale=0.28]{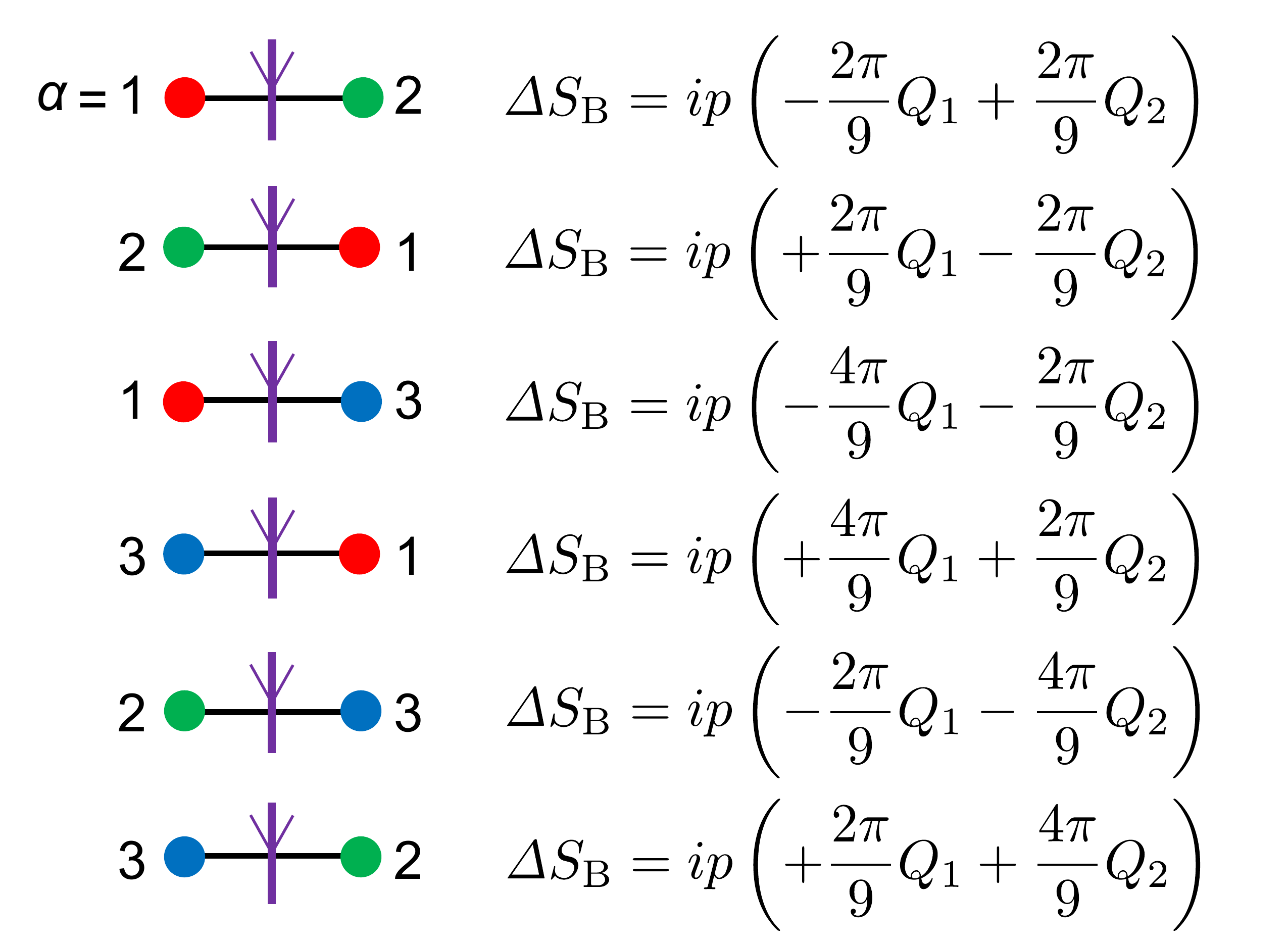}
\caption{Graphical rules for calculating the Berry phase with regards to the tunneling process~\eqref{tunnelingexample}. Each contribution $\mathit{\Delta}S_\mathrm{B}$ is added every time the discontinuity line crosses the correspondent side. }
\label{fig:Berryrule2}
\end{figure}

This observation allows us to find the simple graphical rule to compute the Berry phase: 
If the discontinuity line goes across each edge, we add the contribution $\Delta S_{\mathrm{B}}$ shown in Fig.~\ref{fig:Berryrule2} to $S_\mathrm{B}$, 
\begin{equation}
    S_{\mathrm{B}}=\sum_{\mathrm{edges}}\Delta S_\mathrm{B} \;. 
\end{equation}
Since each edge is shared by two triangles, the result shown in Fig.~\ref{fig:Berryrule2} becomes twice compared with that in Fig.~\ref{fig:newBerry1} and gives $\mathbb{Z}_9$ phases, $z=\exp(2\pi\mathrm{i}/9)$. 
Using the graphical rule given in Fig.~\ref{fig:Berryrule2}, we can show that any deformation of discontinuity line does not change $S_\mathrm{B}$ as long as its endpoints are fixed. 
That is, the Berry phase of the monopole events depends only on the location, at which the skyrmion is created/annihilated.

\subsection{Effective theory from the monopole gas}
\label{ssec:VBS_Lagrangian}

Let us discuss the physics of VBS phase using the monopoles following the idea of Refs.~\cite{Read:1989zz, Read:1989jy, Read:1990zza}. 
We first rewrite the $SU(3)/U(1)^2$ nonlinear sigma model in terms of the linear sigma model, 
\begin{equation}
    \mathcal{L}=\sum_{\alpha}|(\partial_\mu+\mathrm{i} a_{\alpha,\mu})\vec{\phi}_\alpha|^2+{1\over 2e^2}|\diff a_\alpha|^2+V(\vec{\phi}_{1,2,3}) \;.
\end{equation}
Here, $a_{\alpha=1,2,3}$ are $U(1)$ gauge fields with $a_3=-(a_1+a_2)$, and $\vec{\phi}_{\alpha}$ are $\mathbb{C}^3$-valued scalar field \textit{without} orthonormality constrain~\eqref{orthonomal}. 
We instead introduce the potential term, $V(\vec{\phi}_{1,2,3})$, which should be consistent with the symmetry constraint:
\begin{align}
    V(\vec{\phi}_{1,2,3})&=m^2\sum_{\alpha}|\vec{\phi}_\alpha|^2+g_1 \sum_{\alpha}|\vec{\phi}_\alpha|^4+g_2\sum_{\alpha\not=\beta}|\vec{\phi}_\alpha|^2|\vec{\phi}_\beta|^2+g_3\sum_{\alpha\not=\beta}|\vec{\phi}^*_\alpha\cdot \vec{\phi}_\beta|^2\nonumber\\
    &+g_4\mathrm{Re}\bigl(\vec{\phi}_1\cdot (\vec{\phi}_2\times \vec{\phi}_3)\bigr)+g_5\bigl[\mathrm{Im}\bigl(\vec{\phi}_1\cdot (\vec{\phi}_2\times \vec{\phi}_3)\bigr)\bigr]^2 \;. 
\end{align}
Here, $e^2$ is the gauge coupling, $m^2$ is the (mass)$^2$ for $\vec{\phi}_\alpha$ quanta, and $g_{1,2,3,4,5}$ are the scalar coupling constants. 
The mean-field phase diagram of this linear-sigma model can be found in Ref.~\cite{Pimenov:2017nzi}. 
In order to recover the $SU(3)/U(1)^2$ nonlinear sigma model, we can take $m^2<0$, $g_1>0$, $g_2=0$, $g_3>0$, $g_4<0$ and $g_5>0$. 
The $g_3$ term reproduces the orthogonality constraint, and the $g_4$ and $g_5$ terms require that $\vec{\phi}_1\cdot (\vec{\phi}_2\times \vec{\phi}_3)>0$. 
We then find that the classical vacua of $\vec{\phi}_\alpha$ become $SU(3)$ up to the overall normalization, and the low-energy effective theory is given by the $SU(3)/U(1)^2$ nonlinear sigma model. 

When $m^2<0$, the $SU(3)$ symmetric vacuum $\vec{\phi}_\alpha$ is unstable, so $PSU(3)$ symmetry is spontaneously broken to its subgroup, and we can describe various magnetic orders depending on the coupling constaints $g_{1,2,3,4,5}$. 
By considering the opposite case, $m^2>0$, the $SU(3)$ symmetric point, $\vec{\phi}_\alpha=0$, becomes the classical minimum. 
This is a candidate of the theory that can describe the VBS order. 
By integrating out the $\vec{\phi}_\alpha$ fields, we just obtain the pure $U(1)\times U(1)$ gauge theory with the $S_3$ permutation symmetry. 
Using the $3$d Abelian duality, we can rewrite it in terms of the $2\pi$-periodic scalar fields $\sigma_{1,2}$:
\begin{equation}
    \mathcal{L}^{\mathrm{dual}}={e^2\over 24\pi^2}\left(|\diff \sigma_1|^2+|\diff \sigma_2|^2+|\diff \sigma_1-\diff \sigma_2|^2\right) \;.
\end{equation}
For details of the $3$d Abelian duality, see Appendix~\ref{sec:3d_Abelian_duality}. 
Since $\sigma_{1,2}$ are $2\pi$ periodic scalars, the local operators should take the form of 
\begin{equation}
    \exp(-\mathrm{i} Q_1 \sigma_1-\mathrm{i} Q_2 \sigma_2) \;, 
\end{equation}
with $Q_1,Q_2\in \mathbb{Z}$. 
This is the monopole operator with the topological charge $(Q_1,Q_2,-Q_1-Q_2)$. 

\begin{figure} \centering
\includegraphics[scale=0.4]{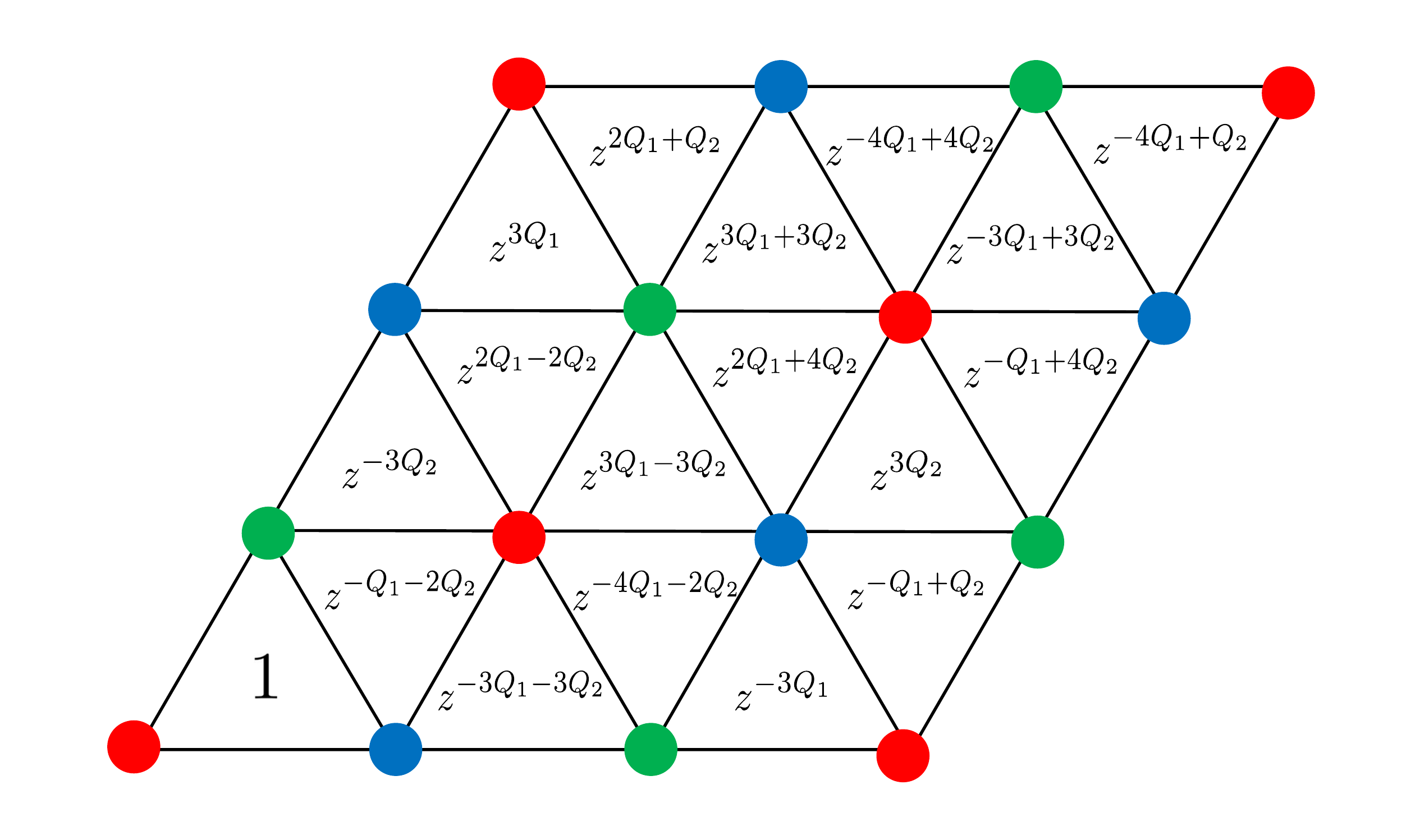}
\caption{Result of the Berry-phase factor, $\exp(-S_{\mathrm{B}})$, for the tunneling process~\eqref{tunnelingexample} when $p=1$. 
Skyrmions are supposed to be created and annihilated at the middle of the lattice. 
We consider the case where the skyrmion is created at the lower left triangle, and the factor in each triangle gives $\exp(-S_{\mathrm{B}})$ when the skyrmion is annihilated there. 
Here, $z=\exp(2\pi \mathrm{i}/9)$ is the $9$th root of unity.  }
\label{fig:Berry2}
\end{figure}

In order to judge if the effective Lagrangian is perturbed by the monopole operator, we need to calculate the \textit{destructive interference} from the Berry phase~\cite{Haldane:1988zz}. 
By use of the graphical rule in Fig.~\ref{fig:Berryrule2}, we can easily calculate Berry phase of specific tunneling processes, as illustrated in Fig.~\ref{fig:Berry2}. 
In Fig.~\ref{fig:Berry2}, the skyrmion is created at the lower left triangle, and we show the phase factor, $\exp(-S_{\mathrm{B}})$, for each triangle, at which the skyrmion is annihilated. 
When the skyrmion is created at other triangles, the phase factors must be divided by the factor of the starting point. 
We have set $p=1$ in this discussion. 
This $18$ consecutive triangles give the minimal unit for the repeated pattern of the Berry phase. 

In order to compute the destructive interference, we have to add these $18$ phase factors, which correspond to the process of the skyrmion annihilation in neighboring triangles, and it gives 
\begin{equation}
    (1+z^{3Q_1}+z^{-3Q_1})(1+z^{3Q_2}+z^{-3Q_2})(1+z^{-Q_1-2Q_2}) \;, 
\end{equation}
with $z=\exp(2\pi \mathrm{i}/9)$.
To generalize the result for $p>1$, we just need to replace $(Q_1,Q_2)\to (pQ_1,pQ_2)$. 
Since $1+z^3+z^{-3}=0$, the monopole operator cannot contribute to the Lagrangian unless $Q_1,Q_2\in 3\mathbb{Z}$. 
This suggests that the topological symmetry $U(1)^2_{\mathrm{top}}$ is explicitly broken to 
\begin{equation}
    (\mathbb{Z}_3)_{\mathrm{top}}^2\subset U(1)_{\mathrm{top}}^2 \;. 
\end{equation}
Although the continuous topological symmetry is gone, we still have a nontrivial subgroup, which plays the important role.

In order to obtain the monopole effective potential, let us use the dilute monopole gas approximation~\cite{Polyakov:1976fu}. The minimal allowed charges are $(Q_1,Q_2,Q_3)=(3,0,-3)$, $(0,3,-3)$, and $(3,-3,0)$, and their charge conjugates. Because of the $1+z^{-Q_1-2Q_2}$ factor in the destructive interference, we find 
\begin{align}
    V_{\mathrm{eff}}&=-\left\{\cos\left(3\sigma_1\right)+\cos\left(3\sigma_1+{2\pi\over 3}\right)\right.\nonumber\\
    &\qquad+\cos\left(3\sigma_2\right)+\cos\left(3\sigma_2-{2\pi\over 3}\right)\nonumber\\
    &\qquad\left.+\cos\left(3(\sigma_1-\sigma_2)\right)+\cos\left(3(\sigma_1-\sigma_2)-{2\pi\over 3}\right)\right\} \;. 
\end{align}
The minima of this potential are given by 
\begin{equation}
    (\sigma_1,\sigma_2)=\left({2\pi\over 3}n_1,{2\pi\over 3}n_2\right), \,\,\left({2\pi\over 3}n_1-{2\pi\over 9},{2\pi\over 3}n_2+{2\pi\over 9}\right) \;, 
\end{equation}
and there are $18$ degenerate vacua. 
Because of the spontaneous breaking of $(\mathbb{Z}_3)_{\mathrm{top}}^2$, the ground-state degeneracy has to be a multiple of $9$, and we obtain the consistent result. 
Extra factor two comes from the fact that one of the $(\mathbb{Z}_3)_{\mathrm{top}}$ symmetry is indeed a subgroup of the $\mathbb{Z}_6$ symmetry, 
\begin{equation}
    (\mathbb{Z}_6)_{\mathrm{top}}:\begin{pmatrix}
    \sigma_1\\
    \sigma_2
    \end{pmatrix}\mapsto 
    \begin{pmatrix}
    1&-1\\
    0&-1
    \end{pmatrix}\begin{pmatrix}
    \sigma_1\\
    \sigma_2
    \end{pmatrix}
    +{2\pi\over 9}\begin{pmatrix}
    -1\\
    1
    \end{pmatrix}, 
\end{equation}
and we may identify it as the $(\mathbb{Z}_6)_{\mathrm{rot}}$ symmetry. 

When $p$ is a multiple of $3$, the story is drastically different. We can see that $U(1)^2_{\mathrm{top}}$ is completely broken explicitly, since the destructive interference does not occur. 
% As a result, we obtain the unique vacuum by the dilute monopole gas approximation when $p\in 3\mathbb{Z}$. 

We would like to make the connection between the effective theory discussed above and the VBS phase of the lattice model, but there is a puzzle. 
When we construct the VBS order on the lattice for $p=1$, we find $6$ degenerate vacua as a consequence of the sponteneous breaking of the lattice rotation symmetry, $(\mathbb{Z}_6)_{\mathrm{rot}}\to 1$. 
On the other hand, our continuum theory has $(\mathbb{Z}_6)_{\mathrm{top}}\times (\mathbb{Z}_3)_{\mathrm{top}}$ and the monopole effective theory has $18$ degenerate vacua due to its symmetry breaking. 
So, there is a mismatch between them, and we currently do not know how we should modify the monopole theory to resolve it. 
We leave this issue as an open problem for future study, and it would be important to analyze the VBS phase using the field-theory language in more details.

%%%%%%%%%%%%%%%%%%%%%%%%%%%%%%%%%%%%%%%%%%
%%%%%%%%%%%%%%%%%%%%%%%%%%%%%%%%%%%%%%%%%%
\section{Phase transition and constraints by anomaly matching }  \label{sec:phase}

In this section, we compute the 't~Hooft anomaly of the $SU(3)/U(1)^2$ sigma model, and discuss the constraint of possible dynamics using anomaly matching condition. 
This is a technique originally developed in the context of chiral symmetry in quantum chromodynamics (QCD)~\cite{tHooft:1979rat, Frishman:1980dq, Coleman:1982yg, Wess:1971yu}, and is analogous to Lieb-Schultz-Mattis theorem in $(1+1)$d spin chains~\cite{Lieb:1961fr, Affleck:1986pq}. 
Thanks to theoretical developments of symmetry-protected topological (SPT) phases, we now obtain much deep understandings of anomaly matching condition~\cite{Wen:2013oza, Cho:2014jfa, Kapustin:2014lwa, Kapustin:2014zva, Wang:2014pma} (see, for example,  Refs.~\cite{Seiberg:2016rsg, Tachikawa:2016cha, Gaiotto:2017yup, Tanizaki:2017bam, Komargodski:2017dmc, Komargodski:2017smk, Shimizu:2017asf, Gaiotto:2017tne, Tanizaki:2017qhf, Tanizaki:2017mtm,Dunne:2018hog, Tanizaki:2018wtg,  Sulejmanpasic:2018upi, Yao:2018kel, Furusawa:2020kro} for some of recent applications). 

We first derive 't~Hooft anomalies for noncompact $SU(3)/U(1)^2$ sigma model in Sec.~\ref{ssec:PSU(3)top} based on Ref.~\cite{Tanizaki:2018xto}. 
Since there are no dynamical monopoles, we can easily find the anomaly involving $SU(3)/\mathbb{Z}_3$ spin symmetry and $U(1)^2$ topological symmetry using the continuum description. 
After that, in Sec.~\ref{sec:monopoleanom}, we take into account the effect of dynamical monopoles, and discuss its implications to the N\'eel-VBS phase transition.

\subsection{'t Hooft anomaly of noncompact $SU(3)/U(1)^2$ sigma model}  \label{ssec:PSU(3)top}

In this subsection, we review the  computation of 't~Hooft anomalies of \textit{noncompact} $SU(3)/U(1)^2$ sigma model based on  Ref.~\cite{Tanizaki:2018xto}. 
Here, the adjective ``noncompact'' is a jargon to imply the absence of dynamical monopoles, so that the theory has the topological symmetry, $U(1)_{\mathrm{top}}^2$. 

We here use the $U(1)^2$ gauge theory description given in Eq.~\eqref{SU(3)effactionA}. 
In order to detect anomalies, we have to perform the gauging of global symmetries discussed in Sec.~\ref{sec:symmetry}. 
Let us first consider the gauging of $PSU(3)$ spin symmetry. 
By regarding $PSU(3)=U(3)/U(1)$, background $PSU(3)$ gauge field consists of $1$-form and $2$-form gauge fields~\cite{Kapustin:2014gua}:\footnote{Here, we denote  background gauge fields by capital letters, while dynamical fields are denoted by lowercase. }
\begin{itemize}
    \item $A$: $U(3)$ $1$-form gauge field, 
    \item $B$: $U(1)$ $2$-form gauge field. 
\end{itemize}
We introduce the following constraint between these two gauge fields,
\begin{equation}
    3B=\diff(\tr(A)) \;. 
\end{equation}
This equation is invariant under the $U(1)$ $1$-form gauge transformation,
\begin{equation}
    B\mapsto B+\diff \xi \;,\quad A\mapsto A+\xi I_3 \;, 
\end{equation}
where the gauge parameter $\xi$ is another $U(1)$ gauge field. 
By requiring this $1$-form gauge invariance at every stage of the gauging procedure, we can correctly gauge $PSU(3)$ symmetry. 

Since $A$ behaves as an $SU(3)$ gauge field locally, it couples to $\vec{\phi}_\alpha$ by the minimal coupling procedure, so we perform the following replacement of the covariant derivative, 
\begin{equation}
    (\diff+\im\, a_\alpha)\vec{\phi}_{\alpha}\to (\diff +\im\, a_\alpha +\im A)\vec{\phi}_\alpha \;. 
\end{equation}
To achieve the $1$-form gauge invariance of the covariant derivative, the $U(1)$ gauge fields should transform as 
\begin{equation}
    a_\alpha\mapsto a_\alpha-\xi \;. 
\end{equation}
Since the constraint~\eqref{eq:constraint_gaugefield} does not satisfy the $1$-form gauge invariance, we should replace it by 
\begin{equation}
    \sum_{\alpha=1}^{3}a_\alpha+\tr(A)=0 \;. 
\end{equation}
In this way, we obtain the $PSU(3)$-gauged action, 
\begin{equation}
    S_{\mathrm{gauged}}=\int_{M_3} \diff^3 x {1\over g_{\mathrm{eff}}}\sum_{\alpha=1}^{3} \Bigl|(\partial_\mu +\im\, a_{\alpha,\mu}+\im A_\mu)\vec{\phi}_\alpha\Bigr|^2 \;, 
    \label{eq:gauged_action_3d}
\end{equation}
where $M_3$ is the $3$d Euclidean spacetime manifold. 
We note that the $U(1)$ field strength, $\diff a_\alpha$, is not invariant under $1$-form gauge transformations, so we have to replace it as 
\begin{equation}
    \diff a_\alpha \to \diff a_\alpha + B \;. 
\end{equation}
This is important when discussing the topological symmetry. 

Let us introduce the background gauge fields for $U(1)^2_{\mathrm{top}}$, and denote them as $A^{\mathrm{top}}_1$ and $A^{\mathrm{top}}_2$. 
As the Noether currents are given by Eq.~\eqref{SU(3)symTopcurrent}, the minimal coupling procedure introduces the additional term to $S_\mathrm{gauged}$,
\begin{equation}
    \int_{M_3}{\im \over 2\pi} \left(A^{\mathrm{top}}_1\wedge (\diff a_1+B)+A^{\mathrm{top}}_2 \wedge (\diff a_2+B)\right) \;. 
    \label{eq:topological_coupling_monopole}
\end{equation}
This is invariant under the $1$-form gauge transformation by construction. 
Under the $U(1)^2_{\mathrm{top}}$ gauge transformation, 
\begin{equation}
    A_1^{\mathrm{top}}\mapsto A_1^{\mathrm{top}}+\diff \lambda_1 \;,\quad 
    A_2^{\mathrm{top}}\mapsto A_2^{\mathrm{top}}+\diff \lambda_2 \;, 
\end{equation}
the gauged action changes by
\begin{align}
    \Delta S_{\mathrm{gauged}}&={\im \over 2\pi}\int_{M_3} \left(\diff \lambda_1 \wedge (\diff a_1+B)+\diff \lambda_2\wedge (\diff a_2 + B)\right) \nonumber\\
    &={\im \over 2\pi}\int_{M_3} (\diff \lambda_1+\diff \lambda_2)\wedge B,\quad  (\bmod\,\, 2\pi \im) \;. 
    \label{eq:anomaly_topologicalsymmetry}
\end{align}
This is the 't~Hooft anomaly involving $PSU(3)$ spin symmetry and $U(1)_{\mathrm{top}}^2$. 
Importantly, the anomaly is described solely by background gauge fields, and a potentially anomalous term including dynamical fields vanishes $\bmod\, 2\pi \im$ due to the Dirac quantization condition. 
This anomaly can be understood using the anomaly inflow from the $(3+1)$d SPT order,
\begin{equation}
    S_{4\mathrm{d\, SPT}}={\im\, k \over 2\pi}\int_{M_4} (\diff A_1^{\mathrm{top}}+\diff A_2^{\mathrm{top}})\wedge B \;, 
    \label{eq:4d_SPT_action}
\end{equation}
with $k=1$. Here, the level $k$ has to be an integer for the large gauge invariance on closed $4$-manifolds $M_4$. 
Furthermore, because of $3B=\diff(\tr(A))$, there is an identification, $k\sim k+3$, i.e., we can trivialize this SPT state by stacking its three copies. 
When $M_4$ has the boundary, $\partial M_4=M_3$, the SPT action~\eqref{eq:4d_SPT_action} has the gauge anomaly at the boundary, which exactly cancels the anomaly~\eqref{eq:anomaly_topologicalsymmetry} of $3$d $SU(3)/U(1)^2$ sigma model when $k=1$. 

We can find more subtle anomaly that involves the $S_3$ permutation symmetry by considering the $S^1$ compactification with nontrivial background holonomy~\cite{Tanizaki:2017qhf, Tanizaki:2017mtm,Dunne:2018hog}. 
In principle, it should be possible to find the anomaly directly working on $3$d spacetimes with various backgrounds. 
However, since $S_3$ permutation does not commute with $U(1)^2_{\mathrm{top}}$, the direct computation in $3$d spacetime requires a careful treatment of various cohomological objects. 
For example, since the cyclic permutation, $(\mathbb{Z}_3)_{\mathrm{trans}}\subset S_3$, acts on $\diff a_1$ and $\diff a_2$ as $\diff a_1\mapsto \diff a_2$ and $\diff a_2\mapsto \diff a_3=-\diff a_1-\diff a_2$, the background gauge fields $A^{\mathrm{top}}_1, A^{\mathrm{top}}_2$ are also transformed:
\begin{equation}
\label{eq:monopole_gauge_field_permutation}
    (\mathbb{Z}_3)_{\mathrm{trans}}:\,
    \begin{pmatrix}
    A^{\mathrm{top}}_1\\
    A^{\mathrm{top}}_2
    \end{pmatrix}
    \mapsto \begin{pmatrix}
    -1& 1\\
    -1&0
    \end{pmatrix}
    \begin{pmatrix}
    A^{\mathrm{top}}_1\\
    A^{\mathrm{top}}_2
    \end{pmatrix} \;. 
\end{equation}
Obviously, this $\mathbb{Z}_3$ transformation does not commute with the $U(1)^2$ gauge transformation of $A^{\mathrm{top}}_{1,2}$. 
Because of this discrete non-Abelian feature, we have to replace the derivatives in $\diff A^{\mathrm{top}}$ by the appropriate ``covariant derivative'', $D A^{\mathrm{top}}$, so that it gives an element of $H^2(B(U(1)^2\rtimes \mathbb{Z}_3), U(1))$. 
In certain cases, we can skip these complications by considering $M_3=S^1\times M_2$ with small $S^1$ and introducing the holonomy background along $S^1$. Here, $M_2$ is some $2$d Euclidean space, which is sufficiently larger than the size of $S^1$. 

Let us derive the $2$d effective theory on $M_2$ with the following holonomy background,
\begin{equation}
    \int_{S^1} A^{\mathrm{top}}_1=\theta_1,\quad 
    \int_{S^2} A^{\mathrm{top}}_2=\theta_2 \;, 
\end{equation}
where $\theta_1,\theta_2$ are constants defined in $\mathbb{R}/2\pi \mathbb{Z}$. 
Performing the naive $2$d reduction of \eqref{eq:gauged_action_3d} and \eqref{eq:topological_coupling_monopole}, we obtain
\begin{align}
    S_{2\mathrm{d\, EFT}}&=\int_{M_2} \diff^2 x {1\over g_{2\mathrm{d}}}\sum_{\alpha=1}^{3} \Bigl|(\partial_\mu +\im\, a_{\alpha,\mu}+\im A_\mu)\vec{\phi}_\alpha\Bigr|^2\nonumber\\
    &+\int_{M_2}{1\over 2\pi}\left(\theta_1 \diff a_1+\theta_2\diff a_2\right) \;,  
    \label{eq:2d_EFT}
\end{align}
where $g_{2\mathrm{d}}=L/g_{\mathrm{eff}}$ and $L$ is the circumference of $S^1$.  
This is exactly the $2$d $SU(3)/U(1)^2$ nonlinear sigma model found in Refs.~\cite{Bykov:2011ai, Bykov:2012am, Lajko:2017wif}. 
When we obtain \eqref{eq:2d_EFT} from the Haldane limit of the $1$d $SU(3)$ antiferromagnetic spin chain, the vacuum angles $\theta_1,\theta_2$ take quantized values, 
\begin{equation}
    \theta_1={2\pi\over 3}n\;,\quad \theta_2=-{2\pi\over 3}n\;, 
\end{equation}
with some $n\in \mathbb{Z}$. This integer $n$ comes from the $n$-box symmetric representation on each site of the $1$d spin chain. 
When $n\not = 0 \bmod 3$, the $SU(3)$ spin chain is subject to the Lieb-Schultz-Mattis theorem~\cite{Lajko:2017wif, Affleck:1986pq}, and thus the ground state cannot be simultaneously gapped and translationally invariant.  
We can reach the same conclusion from the mixed anomaly of $PSU(3)$ spin symmetry and $(\mathbb{Z}_3)_{\mathrm{trans}}$ cyclic permutation~\cite{Tanizaki:2018xto}. 
At generic values of $\theta$ angles, the $(\mathbb{Z}_3)_{\mathrm{trans}}$ symmetry is explicitly broken due to the $\theta$ terms in \eqref{eq:2d_EFT}, but it is a good symmetry for the above values of $\theta$ thanks to the Dirac quantization condition. 
To find the anomaly, let us again introduce the $PSU(3)$ background gauge field $(A,B)$, and then the $(\mathbb{Z}_3)_{\mathrm{trans}}$ permutation transforms the topological term as 
\begin{align}
    &\int_{M_2}{n\over 3}\Bigl((\diff a_1+B)-(\diff a_2+B)\bigr)\nonumber\\
    \mapsto& 
    \int_{M_2}{n\over 3}\Bigl((\diff a_2+B)-(-(\diff a_1+B)-(\diff a_2+B))\Bigr)\nonumber\\
    =&\int_{M_2}{n\over 3}\Bigl((\diff a_1+B)-(\diff a_2+B)\bigr)+n\int_{M_2} B\qquad (\bmod\, 2\pi) \;.
    \label{eq:2d_anomaly}
\end{align}
As long as $n\not=0 \bmod 3$, the last term is nontrivial and gives a mixed 't~Hooft anomaly between $PSU(3)$ and $(\mathbb{Z}_3)_{\mathrm{trans}}$. 

We can now go back to the discussion on $3$d $SU(3)/U(1)^2$ sigma model. In obtaining $2$d theory with 't~Hooft anomaly, we have introduced the holonomy background with $\int_{S^1} A_1=-\int_{S^1}A_2=2\pi/3$. 
Since $A_1+A_2=0$ in this setup, the previous $4$d SPT action~\eqref{eq:4d_SPT_action} becomes trivial. 
Therefore, the presence of the  anomaly~\eqref{eq:2d_anomaly} in $2$d theory shows that the full structure of 't~Hooft anomaly has a more detailed structure than  \eqref{eq:4d_SPT_action}.

\subsection{Effect of dynamical monopoles}  \label{sec:monopoleanom}

As we have seen in Sec.~\ref{ssec:VBS_Lagrangian}, the presence of dynamical monopoles explicitly break the $U(1)^2_{\mathrm{top}}$ symmetry. 
When $p\in 3\mathbb{Z}$, it is completely broken, and the 't~Hooft anomaly discussed above does not exist, and the unique gapped ground state may appear. 

When $p\not=0 \bmod 3$, there is a nontrivial subgroup, $(\mathbb{Z}_3)_{\mathrm{top}}^2\subset U(1)_{\mathrm{top}}^2$, and this is sufficient to have an 't~Hooft anomaly~\cite{Tanizaki:2018xto}. 
In this case, we should regard ${1\over 2\pi}\diff A_i^{\mathrm{top}}$ as the Bockstein element, which gives $\mathbb{Z}_3$ $2$-cocycle, and then the topological action~\eqref{eq:4d_SPT_action} is nontrivial. 
In order to match the 't~Hooft anomaly, the low-energy effective theory must include one of the followings;
\begin{itemize}
    \item spontaneous breaking of $PSU(3)$, 
    \item spontaneous breaking of $ (\mathbb{Z}_3)_{\mathrm{top}}^2\,$, 
    \item gapless excitations without symmetry breaking such as conformal field theory, or 
    \item intrinsic topological order. 
\end{itemize}
When we can assume that the system matches the 't~Hooft anomaly by breaking the global symmetry to its anomaly-free subgroup, then we conclude that there has to be a direct phase transition between the N\'eel phase and the VBS phase. 

Let us also argue that this scenario is very preferred from the consideration of dynamics. 
When the system is in the N\'eel phase, $PSU(3)$ spin symmetry is spontaneously broken by the vacuum expectation value of the spinon field $\langle \vec{\phi}_\alpha\rangle \not = 0$, or by its gauge-invariant combination more rigorously. 
In such a region, the gauge field $a_{\alpha}$ is an auxiliary degree of freedom, and the singularity of its field strength $\diff a_\alpha$ is likely to be suppressed. 
This suggests that the N\'eel phase and the VBS phase will not coexist. 

When $PSU(3)$ symmetry is restored, and the $\vec{\phi}_\alpha$ quanta become massive, the relevant degrees of freedom switches to monopoles. 
As long as the destructive interference due to the Berry phase is circumvented, it is natural to expect that the monopole gas contributes to the partition function, which makes the system gapped by the Polyakov mechanism. 
For $p\not=0 \bmod 3$, there is a nontrivial topological symmetry. 
Monopole gas spontaneously breaks the topological symmetry, and it should be the field-theoretic interpretation of the VBS phase. 
Even though there is a puzzle to make the direct correspondence between the monopole effective theory and the VBS phase of the lattice model (see the last paragraph of Sec.~\ref{ssec:VBS_Lagrangian}), the above discussion suggests that the monopole condensation occurs immediately after the restoration of $PSU(3)$ symmetry. 

What would be the order of the N\'eel-VBS phase transition in this system? 
As long as we consider the noncompact field-theoretic model, the phase transition is controlled by the potential for the spinon field $\vec{\phi}_\alpha$. 
According to the standard mean-field discussion with the Ginzburg-Landau theory, we find the second-order transition between the $PSU(3)$ broken and restored phases. 
Therefore, the question is whether dynamical monopoles can change the order of the phase transition. 
If the allowed monopole operators contain the relevant perturbation, it would make the system gapped, and the phase transition becomes first order. 
If all of them turn out to be irrelevant, the phase transition remains to be of second order, and we will find the new deconfined quantum critical point by the $(2+1)$d $SU(3)$ antiferromagnets.

%%%%%%%%%%%%%%%%%%%%%%%%%%%%%%%%%%%%%%%%%%%%%%%%%%%%%%%%%%%%%%
\section{Summary and discussion}  \label{sec:conclusion}

In this paper, we discussed low-energy behavior of the $SU(3)$ antiferromagnetic triangular lattice with $p$-box symmetric representation. 
This is an analogue of the $SU(2)$ antiferromagnets on the square lattice, and it can be a nice playground to generalize the deconfinend quantum critical point of the N\'eel-VBS phase transition~\cite{Senthil:2003eed, PhysRevB.70.144407}. 
By taking the continuum limit of lattice theory for large $p$, we give an explicit derivation of the relativistic $SU(3)/U(1)^2$ nonlinear sigma model. 
In particular, we carefully deal with the possible topological-like terms, and confirm that they completely vanish as long as the continuum approximation is valid.

In order to treat the N\'eel-VBS phase transition in the field-theoretic technique, it is crucial to understand the properties of monopoles, which gives the tunneling process that changes the skyrmion numbers. 
As the homotopy of the sigma-model target  space is given by $\pi_2(SU(3)/U(1)^2)\simeq \mathbb{Z}^2$, there are two kinds of skyrmion numbers, and this makes the situation more complicated than that of the $\mathbb{C}P^1$ sigma model for $SU(2)$ antiferromagnets.  
For explicit computations of the monopole amplitudes, we constructed a convenient parametrization of $SU(3)/U(1)^2$ to describe the low-energy excitations from the classical N\'eel order beyond the continuum approximation. 
This allows us to have an explicit formula to describe the tunneling process with the jump of topological charges, $(0,0,0)\to (Q_1,Q_2,Q_3)\to (0,0,0)$, with generic $Q_\alpha$ satisfying $Q_1+Q_2+Q_3=0$. 

Having this explicit spacetime configuration with monopoles, we calculate the Berry phase for the monopole amplitudes by evaluating the Wess-Zumino term of the spin coherent path integral. 
Surprisingly, we can find a simple graphical rule to compute the Berry phase, and the interference pattern of monopoles is obtained. 
Especially, when $p\not\in 3\mathbb{Z}$, there is the destructive interference for the monopoles unless $Q_1,Q_2\in 3\mathbb{Z}$, and the possible monopole perturbations are constrained.
That is, the $U(1)^{2}_{\mathrm{top}}$ symmetry of the noncompact $SU(3)/U(1)^2$ model is explicitly broken to its nontrivial subgroup, $(\mathbb{Z}_3)_{\mathrm{top}}^2$. 
We note that one of the $(\mathbb{Z}_3)_{\mathrm{top}}$ symmetry is embedded into the rotational symmetry, $(\mathbb{Z}_6)_{\mathrm{top}}\simeq (\mathbb{Z}_6)_{\mathrm{rot}}$. 
When monopole condensation occurs, this topological symmetry is spontaneously broken, which leads to $18$ degenerate ground states, and we identify it as a signature of the VBS phase. 

We compute the 't~Hooft anomaly for this system, and find that there is a nontrivial anomaly when $p\not\in 3\mathbb{Z}$. 
As a consequence of the anomaly matching, low-energy effective theory must contain a nontrivial field theory, and the direct phase transition between the N\'eel and VBS phases is strongly supported. 
We also argued that this is a natural scenario to match the 't~Hooft anomaly for this system from the consideration of dynamics. 

There are many interesting problems that need to be understood in future study. 
One of them, which would be most important, is to resolve the mismatch of the ground-state degeneracies between the monopole effective theory and the VBS orders in the lattice model with $p=1$. 
According to the monopole effective theory constructed in this paper, the $(\mathbb{Z}_6)_{\mathrm{top}}\times (\mathbb{Z}_3)_{\mathrm{top}}$ symmetry is spontaneously broken completely, and we conclude $18$ degenerate vacua. 
On the other hand, in the lattice model on the triangular lattice, the VBS phase breaks the lattice rotation symmetry, $(\mathbb{Z}_6)_{\mathrm{rot}}\to 1$, and has $6$ degenerate ground states. 
We obviously need a better understanding on how the lattice symmetry acts on monopoles to have the complete picture of the VBS phase. 

It is also interesting to ask if we can create the intrinsic topological order by deformation of the $SU(3)$ antiferromagnets. 
From considerations on dynamics with kinematical constraints, we have argued that the N\'eel-VBS direct phase transition is very natural. 
However, as a logical possibility from anomaly matching, the symmetric topological order can match the 't~Hooft anomaly so it can appear as the ground state of some local Hamiltonian with the same symmetry. 

As this study is motivated by the deconfined quantum critical ponit of the $SU(2)$ antiferromagnets, we are also curious about the order of the phase transition between the N\'eel and VBS phases. 
When we neglect the effect of dynamical monopoles, the phase transition is controlled by the change of the (mass)$^2$ term of the spinon field, $\vec{\phi}_\alpha$, and it should be the second-order phase transition. 
We have to understand if the dynamical monopoles are relevant or irrelevant at this fixed point in order to judge whether the massive deformation occurs or not. 
There are many numerical analysis for the case of $SU(2)$ antiferromagnets (see, for exmaple, Refs.~\cite{Sandvik:2006fpf,Sandvik:2010ag,Nahum:2015jya,Nahum:2015vka,Kuklov:2008zz,Bartosch:2013vma,Sreejith:2015ria}). 
Although the problem seems not to be settled yet, it is suggested that the N\'eel-VBS transition is weak first order or second order. 
This problem leads to the development of many infrared duality for the $3$d field theories~\cite{Son:2015xqa,Mross:2015idy,Xu:2015lxa,Seiberg:2016gmd,Wang:2017txt}, and the generalizations and applications of these techniques to the $SU(3)/U(1)^2$ sigma model would be an interesting future study.

%%%%%%%%%%%%%%%%%%%%%%%%%%%%%%%%%%%%%%%%%%%%%%%
\begin{acknowledgments}
Y.~T. appreciates useful discussions in the workshop at Yukawa Institute for Theoretical Phyiscs, ``Topological Phase and Quantum Anomaly 2021'' (YITP-T-21-03). 
The work of Y.~T. was partially supported by JSPS KAKENHI
Grant-in-Aid for Research Activity Start-up, 20K22350.
\end{acknowledgments}

%%%%%%%%%%%%%%%%%%%%%%%%%%%%%%%%%%%%%%%%%%%%%%%
\appendix

\section{Details of the semiclassical derivation of nonlinear sigma model} \label{sec:A}

Appendix~\ref{sec:A} is a supplement for Sec.~\ref{sec:NLS}. 
The formulas~\eqref{coherent1}--\eqref{coherent4} for the spin coherent states are proven in Appendix~\ref{ssec:spin_coherent}. 
The full derivation of the $SU(3)/U(1)^2$ nonlinear sigma model is given in Appendix~\ref{ssec:apSU(3)NLS}. 
The following Appendix~\ref{ssec:apSU(3)top} is devoted to the proof that the topological terms~\eqref{SU(3)efftop} vanish. 

\subsection{Derivation of formulas for the spin-coherent states}\label{ssec:spin_coherent}

Our derivation of the nonlinear sigma model crucially depends on the spin coherent state $|\vec{\Phi}\rangle$, which is defined through the Schwinger boson by the formula~\eqref{coherentdif}. 
In this subsection, we give the proof of its properties~\eqref{coherent1}--\eqref{coherent4}. 

Let us first show the property~\eqref{coherent3} about the inner product; 
\begin{align}
    \langle \vec{\Phi'}|\vec{\Phi}\rangle&={1\over p!}\langle 0|(\vec{\Phi}'^*\cdot \vec{\hat{a}})^p (\vec{\Phi}\cdot \vec{\hat{a}}^\dagger)^p | 0\rangle\nonumber\\
    &=\sum_{\ell_1+\cdots+\ell_N=p}{p!\over \ell_1!\cdots \ell_N!}(\Phi_1'^*\Phi_1)^{\ell_1}\cdots (\Phi_N'^*\Phi_N)^{\ell_N}\nonumber\\
    &=(\vec{\Phi}'^*\cdot \vec{\Phi})^p \:. 
\end{align}    
Here, we expanded $(\vec{\Phi}'^*\cdot \vec{\hat{a}})^p$ and  $(\vec{\Phi}\cdot \vec{\hat{a}}^\dagger)^p$ in components and used the Wick contraction to find the result. 
We note that the normalization condition $|\vec{\Phi}|=1$ is not used in the above derivation, so the formula~\eqref{coherent3} is true for any $\vec{\Phi},\vec{\Phi}'\in \mathbb{C}^N$. 
We can obtain \eqref{coherent1} from this formula by putting $\vec{\Phi}'=\vec{\Phi}$ and using $|\vec{\Phi}|=1$. 

Next, let us show the formula~\eqref{coherent4} for the spin operator. We first rewrite it as 
\begin{equation}
    \hat{S}_{vw}=\hat{a}^\dagger_v \hat{a}_w=-\delta_{vw}+\hat{a}_w \hat{a}^\dagger_v \;. 
\end{equation}
Therefore, we find 
\begin{align}
    \langle \vec{\Phi}|\hat{S}_{vw}|\vec{\Phi}\rangle&=-\delta_{vw}+\langle \vec{\Phi}|\hat{a}_w \hat{a}^\dagger_v|\vec{\Phi}\rangle\nonumber\\
    &=-\delta_{vw}+{1\over p!}\langle 0|(\vec{\Phi}^*\cdot \vec{\hat{a}})^p \hat{a}_w \hat{a}^\dagger_v (\vec{\Phi}\cdot \vec{\hat{a}}^\dagger)^p| 0\rangle\nonumber\\
    &=-\delta_{vw}+{1\over p!(p+1)^2}{\partial\over \partial \Phi^*_w}{\partial\over \partial \Phi_v}\left.\langle 0|(\vec{\Phi}^*\cdot \vec{\hat{a}})^{p+1} (\vec{\Phi}\cdot \vec{\hat{a}}^\dagger)^{p+1}| 0\rangle\right|_{|\vec{\Phi}|=1} \;.
\end{align}
In the last expression, we temporarily regard $\vec{\Phi}\in \mathbb{C}^N$ without any constraints when taking the partial derivatives, and put the constraint $|\vec{\Phi}|=1$ after those manipulations. 
By using \eqref{coherent3}, we obtain 
\begin{align}
    \langle \vec{\Phi}|\hat{S}_{vw}|\vec{\Phi}\rangle
    &=-\delta_{vw}+{1\over (p+1)}{\partial\over \partial \Phi^*_w}{\partial\over \partial \Phi_v}\left.(\vec{\Phi}^*\cdot \vec{\Phi})^{p+1}\right|_{|\vec{\Phi}|=1}\nonumber\\
    &=p\,\Phi^*_v \Phi_w \;. 
\end{align}

Lastly, we prove the completeness relation~\eqref{coherent2}. 
Let us introduce the notation,
\begin{equation}
    \hat{I}=\int\diff \Omega_{\Phi} |\vec{\Phi}\rangle \langle \vec{\Phi}| \;, 
\end{equation}
where $\diff \Omega_\Phi$ is the $SU(N)$-invariant measure. 
This is a linear map from the vector space of $p$-box symmetric representation to itself. 
We note that $\hat{I}$ commutes with any $SU(N)$ rotations: Let $\hat{U}$ be the unitary operator, $\hat{U}\cdot |\vec{\Phi}\rangle=|U\vec{\Phi}\rangle$ for each $U\in SU(N)$, then 
\begin{equation}
    \hat{U}\hat{I}\hat{U}^\dagger
    =\int\diff \Omega_{\Phi} |U\vec{\Phi}\rangle \langle U\vec{\Phi}|=\hat{I} \;. 
\end{equation}
By using Schur's lemma, $\hat{I}$ has to be proportional to the identity $1$. By choosing the suitable normalization of the measure, we find \eqref{coherent2}.

\subsection{Computing the continuum Lagrangian}  \label{ssec:apSU(3)NLS} 

In this appendix, we derive the continuum effective Lagrangian~\eqref{SU(3)effaction} from the lattice model. After relabeling the lattice sites $i\leftrightarrow (j,k,\alpha)$ following Fig.~\ref{fig:SU(3)coordinate}, we have introduced the notation $\vec{\Phi}_{\alpha}(j,k)=\vec{\Phi}(i)$. 
In this notation, the Euclidean action~\eqref{action} becomes
\begin{align}
S &= \int_0^\beta d\tau\left[
p\sum_{j,k,\alpha}\vec{\Phi}_{\alpha}^\ast(j,k)\cdot\partial_\tau\vec{\Phi}_{\alpha}(j,k) + Jp^2\sum_{j,k}\left\{ 
\left|\vec{\Phi}_1^\ast(j,k)\cdot\vec{\Phi}_2(j,k)\right|^2 \right. \right. \nonumber \\
&\qquad\qquad + \left|\vec{\Phi}_2^\ast(j,k)\cdot\vec{\Phi}_3(j,k)\right|^2 + \left|\vec{\Phi}_3^\ast(j,k)\cdot\vec{\Phi}_1(j,k)\right|^2 \nonumber \\
&\qquad\qquad + \left|\vec{\Phi}_1^\ast(j,k)\cdot\vec{\Phi}_2(j+1,k-2)\right|^2 + \left|\vec{\Phi}_3^\ast(j,k)\cdot\vec{\Phi}_2(j+1,k-2)\right|^2 \nonumber \\
&\qquad\qquad + \left|\vec{\Phi}_1^\ast(j,k)\cdot\vec{\Phi}_3(j-1,k+1)\right|^2 + \left|\vec{\Phi}_2^\ast(j,k)\cdot\vec{\Phi}_3(j-1,k+1)\right|^2 \nonumber \\
&\qquad\qquad \left.\left. + \left|\vec{\Phi}_2^\ast(j,k)\cdot\vec{\Phi}_1(j,k+1)\right|^2 + \left|\vec{\Phi}_3^\ast(j,k)\cdot\vec{\Phi}_1(j,k+1)\right|^2 
  \right\} \right] .  \label{a:SU(3)action}
\end{align}
In the following computations, it is convenient to introduce $3\times 3$ matrices, 
% \begin{equation}  \label{a:SU(3)trSS}
% \mathcal{S}_\alpha(j,k)=p\,\vec{\Phi}^\ast(j,k)\,\vec{\Phi}^\mathrm{T}(j,k)=p\,U^\dagger(j,k) L^\dagger(j,k) \Lambda_\alpha L(j,k) U(j,k) \;, 
% \end{equation}
\begin{equation}  \label{a:SU(3)Lambda}
\Lambda_1=\left( 
\begin{array}{ccc}
1 & 0 & 0 \\ 0 & 0 & 0 \\ 0 & 0 & 0 
\end{array} \right) \;,\quad 
\Lambda_2=\left( 
\begin{array}{ccc}
0 & 0 & 0 \\ 0 & 1 & 0 \\ 0 & 0 & 0 
\end{array} \right) \;,\quad 
\Lambda_3=\left( 
\begin{array}{ccc}
0 & 0 & 0 \\ 0 & 0 & 0 \\ 0 & 0 & 1 
\end{array} \right) \;. 
\end{equation}
Using these matrices, we can rewrite each term of (\ref{a:SU(3)action}) as 
\begin{equation}
\left|\vec{\Phi}_\alpha^\ast(j,k)\cdot\vec{\Phi}_\beta(j',k')\right|^2=\tr\left(\Lambda_\beta\, LU(j',k')\, U^\dagger L^\dagger(j,k)\,\Lambda_\alpha\, LU(j,k) \,U^\dagger L^\dagger(j',k') \right), 
\end{equation}
where $L(j,k)$ and $U(j,k)$ are defined in \eqref{SU(3)fluc}. 

Next, we expand \eqref{a:SU(3)action} in terms of the lattice constant $a$,  
and consider the continuum approximation by regarding the lattice constant $a$ as a small parameter:
\begin{equation}  \label{a:SU(3)continuum}
(j,k) \;\longmapsto\; (x,y) \simeq \left(3aj+\frac{3}{2}ak,\frac{\sqrt{3}}{2}ak\right) ,\quad 
\sum_{j,k} \;\longmapsto\; \frac{2}{3\sqrt{3}a^2}\int \diff x\diff y \;. 
\end{equation}
It turns out that $L_{ab}(j,k)$ has the mass term at the lattice scale, so we can neglect their derivatives. Therefore,  $L_{12}(j\pm1, k\pm1,\alpha)$ can be approximated by $L_{12}(j,k)= L_{12}(x,y)$. 
On the other hand, $U$ becomes massless at the classical level, so its derivatives play the important role. Therefore, for example, $U(j-1,k+1)$ should be related to the fields at $(j,k)$ by the Taylor expansion as follows,
\begin{align}
U(j-1,k+1) &= U\left(x-{3\over 2}a,y+{\sqrt{3}\over 2}a\right)  \nonumber \\
&= \left[ U-{3\over 2}a\delx U+{\sqrt{3}\over 2}a\dely U+\frac{(3a/2)^2}{2}\delx^2 U-{3\sqrt{3}\over 4}a^2\delx\dely U+\frac{(\sqrt{3}a/2)^2}{2}\dely^2 U \right]\!(x,y)+\odre{a^3} \;.   \label{a:Utaylor}
\end{align}
With some computations, \eqref{a:SU(3)action} becomes 
\begin{align}
S &= \int_0^\beta\! \diff \tau\!\int\! \diff x\diff y \:{2\over 3\sqrt{3}}\Bigl[ \:
12J\left( |L_{12}|^2+|L_{13}|^2+|L_{23}|^2 \right)    \nonumber \\
&\qquad +\frac{2}{a}\left\{ L_{12}(\delt UU^\dagger)_{21}+L_{13}(\delt UU^\dagger)_{31}+L_{23}(\delt UU^\dagger)_{32} \right. \nonumber \\
&\qquad\qquad\qquad\left. +L_{12}^\ast(\delt UU^\dagger)_{12}+L_{13}^\ast(\delt UU^\dagger)_{13}+L_{23}^\ast(\delt UU^\dagger)_{23} \right\}  \nonumber \\
&\qquad +Jp\left\{ -L_{12}\Bigl( 3(\delx UU^\dagger)_{21}+3\sqrt{3}(\dely UU^\dagger)_{21} \Bigr) +L_{12}^\ast\Bigl( 3(\delx UU^\dagger)_{12}+3\sqrt{3}(\dely UU^\dagger)_{12} \Bigr)  \right. \nonumber \\
&\qquad\qquad\qquad -L_{13}\cdot6(\delx UU^\dagger)_{31}+L_{13}^\ast\cdot6(\delx UU^\dagger)_{13} \nonumber \\
&\qquad\qquad\qquad\left. -L_{23}\Bigl( 3(\delx UU^\dagger)_{32}-3\sqrt{3}(\dely UU^\dagger)_{32} \Bigr) +L_{23}^\ast\Bigl( 3(\delx UU^\dagger)_{23}-3\sqrt{3}(\dely UU^\dagger)_{23} \Bigr) \right\} \nonumber \\
&\qquad +\frac{9Jp^2}{4}\left\{ |(\delx UU^\dagger)_{12}|^2+2|(\delx UU^\dagger)_{13}|^2+|(\delx UU^\dagger)_{23}|^2 \right\}  \nonumber \\
&\qquad +\frac{3\sqrt{3}Jp^2}{4}\left\{ (\delx UU^\dagger)_{32}(\dely UU^\dagger)_{23}+(\delx UU^\dagger)_{23}(\dely UU^\dagger)_{32} \right.  \nonumber \\
&\qquad\qquad\qquad\qquad\left. -(\delx UU^\dagger)_{12}(\dely UU^\dagger)_{21}-(\delx UU^\dagger)_{21}(\dely UU^\dagger)_{12} \right\}  \nonumber \\
&\qquad\left. +\frac{15Jp^2}{4}\left\{ |(\dely UU^\dagger)_{12}|^2+\frac{2}{5}|(\dely UU^\dagger)_{13}|^2+|(\dely UU^\dagger)_{23}|^2 \right\} 
\:\right] . 
% S &= \int_0^\beta\! \diff \tau\!\int\! \diff x\diff y \:\left[ \:
% \frac{8J}{\sqrt{3}}\left( |L_{12}|^2+|L_{13}|^2+|L_{23}|^2 \right)  \right.  \nonumber \\
% &\qquad +\frac{4}{3\sqrt{3}a}\left\{ L_{12}(\delt UU^\dagger)_{21}+L_{13}(\delt UU^\dagger)_{31}+L_{23}(\delt UU^\dagger)_{32} \right. \nonumber \\
% &\qquad\qquad\left. +L_{12}^\ast(\delt UU^\dagger)_{12}+L_{13}^\ast(\delt UU^\dagger)_{13}+L_{23}^\ast(\delt UU^\dagger)_{23} \right\}  \nonumber \\
% &\qquad +\frac{2Jp}{3\sqrt{3}}\left\{ -L_{12}\Bigl( 3(\delx UU^\dagger)_{21}+3\sqrt{3}(\dely UU^\dagger)_{21} \Bigr) +L_{12}^\ast\Bigl( 3(\delx UU^\dagger)_{12}+3\sqrt{3}(\dely UU^\dagger)_{12} \Bigr)  \right. \nonumber \\
% &\qquad\qquad -L_{13}\cdot6(\delx UU^\dagger)_{31}+L_{13}^\ast\cdot6(\delx UU^\dagger)_{13} \nonumber \\
% &\qquad\qquad\left. -L_{23}\Bigl( 3(\delx UU^\dagger)_{32}-3\sqrt{3}(\dely UU^\dagger)_{32} \Bigr) +L_{23}^\ast\Bigl( 3(\delx UU^\dagger)_{23}-3\sqrt{3}(\dely UU^\dagger)_{23} \Bigr) \right\} \nonumber \\
% &\qquad +\frac{\sqrt{3}Jp^2}{2}\left\{ |(\delx UU^\dagger)_{12}|^2+2|(\delx UU^\dagger)_{13}|^2+|(\delx UU^\dagger)_{23}|^2 \right\}  \nonumber \\
% &\qquad +\frac{Jp^2}{2}\left\{ (\delx UU^\dagger)_{32}(\dely UU^\dagger)_{23}+(\delx UU^\dagger)_{23}(\dely UU^\dagger)_{32} \right.  \nonumber \\
% &\qquad\qquad\left. -(\delx UU^\dagger)_{12}(\dely UU^\dagger)_{21}-(\delx UU^\dagger)_{21}(\dely UU^\dagger)_{12} \right\}  \nonumber \\
% &\qquad\left. +\frac{5Jp^2}{2\sqrt{3}}\left\{ |(\dely UU^\dagger)_{12}|^2+\frac{2}{5}|(\dely UU^\dagger)_{13}|^2+|(\dely UU^\dagger)_{23}|^2 \right\} 
% \:\right] . 
\label{a:SU(3)lagrangian1}
\end{align}
Here, we truncated the derivative expansion within $O(a^2)$. $(\p_\mu UU^\dagger)_{\alpha \beta}$ is the $(\alpha,\beta)$ component of the $3\times 3$ matrix $\p_\mu U U^\dagger$, which is the unique nonzero component of $\Lambda_{\alpha}\p_\mu U U^\dagger\Lambda_{\beta}$. 
To obtain this expression~(\ref{a:SU(3)lagrangian1}), we used the following identity which follows from the unitarity condition, $U U^\dagger=I_3$: 
\begin{equation}  \label{a:idU1}
\partial_\mu UU^\dagger+U\partial_\mu U^\dagger=0 \;. 
\end{equation}

The first line of \eqref{a:SU(3)lagrangian1} is the mass term for $L_{12}, L_{23}, L_{31}$, which shows that they are high-energy fluctuations as we have expected. 
We integrate out these fast variables $L_{12}$, $L_{13}$ and $L_{23}$ by using Gaussian integration: 
\begin{equation}  \label{a:gauss}
\int \diff z\diff z^\ast\mathrm{e}^{-(z^\ast wz-u^\ast z-vz^\ast)}\propto \mathrm{e}^{u^\ast v/w}. 
\end{equation}
The resultant effective action is\footnote{We  interpret subscripts as modulo $N\,(=3)\,$; e.g. $\vec{\phi}_{N+1}$ means $\vec{\phi}_1$, and $\vec{\phi}_{0}$ means $\vec{\phi}_N$. }
\begin{align}
S_\mathrm{eff} &= \int_0^\beta\! \diff \tau\!\int\! \diff x\diff y \:
\frac{1}{g_{\mathrm{eff}}}\sum_{\alpha=1}^{3}\left\{ \frac{1}{v}\left|(\partial_\tau UU^\dagger)_{\alpha,\alpha+1}\right|^2+v\left( \left|(\partial_x UU^\dagger)_{\alpha,\alpha+1}\right|^2+\left|(\partial_y UU^\dagger)_{\alpha,\alpha+1}\right|^2 \right) \right\}+S_{\mathrm{add}},
\label{a:SU(3)lagrangian2} 
\end{align}
where we defined
\begin{equation}  \label{a:SU(3)gv}
g_{\mathrm{eff}}=\frac{3\sqrt{3}}{\sqrt{2}p}a \:, \quad v=\frac{3}{\sqrt{2}}\,Jap \;. 
\end{equation}
When substituting $p=1$, this matches with the result of Ref.~\cite{PhysRevB.88.184430}. 
We can eliminate $v$ from the effective action~\eqref{a:SU(3)lagrangian2} by rescaling the imaginary time, 
\begin{equation}
    \tau \mapsto {\tau\over v} \;. 
\end{equation}
As a result, we can simply set $v=1$ in \eqref{a:SU(3)lagrangian2} by the redefinition of the inverse temperature as $\beta'= v\beta$. 
The last term, $S_{\mathrm{add}}$, is the additional topological-like term, 
\begin{align}
S_{\mathrm{add}}&=\int\diff^3 x\,\frac{p}{3\sqrt{3}a}\biggl( \Bigl\{ (\partial_x UU^\dagger)_{12}{(\partial_\tau UU^\dagger)_{12}}^{\hspace{-8pt}\ast} \;- (\partial_\tau UU^\dagger)_{12}{(\partial_x UU^\dagger)_{12}}^{\hspace{-8pt}\ast} \:\Bigr\}  \nonumber \\
&\qquad\qquad + \sqrt{3}\Bigl\{ (\partial_y UU^\dagger)_{12}{(\partial_\tau UU^\dagger)_{12}}^{\hspace{-8pt}\ast} \;- (\partial_\tau UU^\dagger)_{12}{(\partial_y UU^\dagger)_{12}}^{\hspace{-8pt}\ast} \:\Bigr\}  \nonumber \\
&\qquad\qquad + 2\Bigl\{ (\partial_x UU^\dagger)_{13}{(\partial_\tau UU^\dagger)_{13}}^{\hspace{-8pt}\ast} \;- (\partial_\tau UU^\dagger)_{13}{(\partial_x UU^\dagger)_{13}}^{\hspace{-8pt}\ast} \:\Bigr\}  \nonumber \\
&\qquad\qquad + \Bigl\{ (\partial_x UU^\dagger)_{23}{(\partial_\tau UU^\dagger)_{23}}^{\hspace{-8pt}\ast} \;- (\partial_\tau UU^\dagger)_{23}{(\partial_x UU^\dagger)_{23}}^{\hspace{-8pt}\ast} \:\Bigr\}  \nonumber \\
&\qquad\qquad  - \sqrt{3}\Bigl\{ (\partial_y UU^\dagger)_{23}{(\partial_\tau UU^\dagger)_{23}}^{\hspace{-8pt}\ast} \;- (\partial_\tau UU^\dagger)_{23}{(\partial_y UU^\dagger)_{23}}^{\hspace{-8pt}\ast} \:\Bigr\}  \biggr) \;. 
\end{align}

The first term of \eqref{a:SU(3)lagrangian2} gives \eqref{SU(3)effaction}. 
This can be seen explicitly after the following manipulations:
\begin{align}
2\sum_{\alpha=1}^3\left|(\delt UU^\dagger)_{\alpha,\alpha+1}\right|^2 
% &= 2\sum_{\alpha=1}^3(\delt\vec{\phi}_{\alpha,\alpha+1}\cdot\vec{\phi}_\alpha^\ast)(\delt\vec{\phi}_{\alpha,\alpha+1}^\ast\cdot\vec{\phi}_\alpha) \nonumber \\
&= -2\sum_{\alpha=1}^3\mathrm{tr}\bigl[ \Lambda_{\alpha+1}U\delt U^\dagger\Lambda_\alpha U\delt U^\dagger \bigr] \nonumber \\
&= -\sum_{\alpha=1}^3\sum_{\alpha'(\neq\alpha)}\mathrm{tr}\bigl[ \Lambda_{\alpha}U\delt U^\dagger\Lambda_{\alpha'} U\delt U^\dagger \bigr] \nonumber \\
&= -\sum_{\alpha=1}^3\mathrm{tr}\bigl[ \Lambda_{\alpha}U\delt U^\dagger(I_3-\Lambda_{\alpha}) U\delt U^\dagger \bigr] \nonumber \\
%&= \sum_{\alpha=1}^3 \mathrm{tr}\bigl[ \Lambda_\alpha(-\delt UU^\dagger)U\delt U^\dagger - \Lambda_\alpha U\delt U^\dagger\Lambda_\alpha U\delt U^\dagger \bigr] \nonumber \\
% &= \sum_{\alpha=1}^3\left( \mathrm{tr}\bigl[ \Lambda_\alpha\delt U\delt U^\dagger \bigr] + \mathrm{tr}\bigl[ \Lambda_\alpha U\delt U^\dagger\Lambda_\alpha U\delt U^\dagger \bigr] \right)  \nonumber \\
&= \sum_{\alpha=1}^3\left( \left|\delt \vec{\phi}_\alpha\right|^2-\left|\vec{\phi}_\alpha^\ast\cdot\delt \vec{\phi}_\alpha\right|^2 \right) \;. \label{a:SU(3)Uphikin}
\end{align}
The same is true for the spatial derivatives, and we obtain \eqref{SU(3)effaction}. 

We can rewrite the topological-like term, $S_{\mathrm{add}}$, using the Chern numbers. The Chern numbers $Q_\alpha^{\mu\nu}\,$, which is defined in \eqref{Qdef}, can be decomposed to
\begin{equation}  \label{a:Qredef}
Q_\alpha^{\mu\nu}=\sum_{\alpha'\,(\neq\alpha)}q_{\alpha\alpha'}^{\mu\nu} \:, 
\end{equation}
where 
\begin{equation}  \label{a:smallq}
q_{\alpha\alpha'}^{\mu\nu}=\frac{-1}{2\pi \im}\int\! \diff x^\mu \diff x^\nu\left\{ \mathrm{tr}\bigl[ \Lambda_\alpha\partial_\mu UU^\dagger\Lambda_{\alpha'}\partial_\nu UU^\dagger \bigr]-\mathrm{tr}\bigl[ \Lambda_{\alpha'}\partial_\nu UU^\dagger\Lambda_\alpha\partial_\mu UU^\dagger \bigr] \right\} . 
\end{equation}
Here, we do not take the summation over the indices $\mu,\nu$. 
By definition, $q_{\alpha\alpha'}^{\mu\nu}$ is asymmetric about $\alpha$ and $\alpha'$. Then, we obtain the expression~\eqref{SU(3)efftop} as follows:
\begin{align}
S_\mathrm{add} &= 2\pi \im\frac{p}{3\sqrt{3}a} \left[ \;
\int \diff y\,q_{12}^{x\tau}+\sqrt{3}\int \diff x\,q_{12}^{y\tau} +2\int \diff y\,q_{13}^{x\tau}+\int \diff y\,q_{23}^{x\tau}-\sqrt{3}\int \diff x\,q_{23}^{y\tau}  \right]   \nonumber \\
&= 2\pi \im\frac{p}{3\sqrt{3}a} \left[
\int \diff y\Bigl(q_{12}^{x\tau}+q_{13}^{x\tau}\Bigr)-\sqrt{3}\int \diff x\Bigl(q_{21}^{x\tau}+q_{23}^{x\tau}\Bigr)-\int \diff y\Bigl(q_{31}^{x\tau}+q_{32}^{x\tau}\Bigr) \right]  \nonumber \\
&= \frac{2\pi \im \,p}{3a}\left[ 
\frac{1}{\sqrt{3}}\int \diff y\,Q_1^{x\tau}(y)-\int \diff x\,Q_2^{y\tau}(x)-\frac{1}{\sqrt{3}}\int \diff y\,Q_3^{x\tau}(y) \right] \;. \label{a:SU(3)Qqtop}
\end{align}

\subsection{Topological-like additional terms}  \label{ssec:apSU(3)top}

In order to show that the topological-like additional terms~\eqref{SU(3)efftop} vanish, we define new axes $x', y'$ along the axes $j, k$ as 
\begin{equation}  \label{a:SU(3)jkprime}
(x',y')=(x-\sqrt{3}y,2y) \;. 
\end{equation}
The topological-like additional term becomes
\begin{align}
S_\mathrm{add} &=  \frac{2\pi\im\, p}{3a}\int_0^\beta \diff \tau\!\int\! \diff x\diff y \frac{1}{2\pi \im}\left[ 
\frac{1}{\sqrt{3}}\left\{\partial_x\vec{\phi}_1\cdot\partial_\tau\vec{\phi}_1^\ast-\partial_\tau\vec{\phi}_1\cdot\partial_x\vec{\phi}_1^\ast\right\}  \right. \nonumber \\
&\qquad\left.  -\left\{\partial_y\vec{\phi}_2\cdot\partial_\tau\vec{\phi}_2^\ast-\partial_\tau\vec{\phi}_2\cdot\partial_y\vec{\phi}_2^\ast\right\}  -\frac{1}{\sqrt{3}}\left\{\partial_x\vec{\phi}_3\cdot\partial_\tau\vec{\phi}_3^\ast-\partial_\tau\vec{\phi}_3\cdot\partial_x\vec{\phi}_3^\ast\right\}  \right]  \nonumber \\
&= \frac{2\pi\im\, p}{3a}\int_0^\beta \diff \tau\!\int\! \frac{1}{2}\diff x'\diff y' \frac{1}{2\pi \im}\left[ 
\frac{1}{\sqrt{3}}\left\{\partial_{x'}\vec{\phi}_1\cdot\partial_\tau\vec{\phi}_1^\ast-\partial_\tau\vec{\phi}_1\cdot\partial_{x'}\vec{\phi}_1^\ast\right\}  \right. \nonumber \\
&\qquad\left.  -\left\{\partial_{y'}\vec{\phi}_2\cdot\partial_\tau\vec{\phi}_2^\ast-\partial_\tau\vec{\phi}_2\cdot\partial_{y'}\vec{\phi}_2^\ast\right\}  -\frac{1}{\sqrt{3}}\left\{(-\sqrt{3}\partial_{x'}+2\partial_{y'})\vec{\phi}_3\cdot\partial_\tau\vec{\phi}_3^\ast-\partial_\tau\vec{\phi}_3\cdot(-\sqrt{3}\partial_{x'}+2\partial_{y'})\vec{\phi}_3^\ast\right\}  \right]  \nonumber \\
&= \frac{2\pi\im\, p}{3a}\left[ \frac{1}{2\sqrt{3}}\int \diff y'Q_1^{x'\tau}(y')-\frac{1}{2\sqrt{3}}\int \diff y'Q_3^{x'\tau}(y')+\frac{\sqrt{3}}{2}\int \diff y'Q_2^{x'\tau}(y')-\int \diff x'Q_2^{y'\tau}(x') \right]  \nonumber \\
&= -\frac{2\pi\im\, p}{3a}\left[ \frac{1}{\sqrt{3}}\int \diff y'Q_1^{x'\tau}(y')+\frac{2}{\sqrt{3}}\int \diff y'Q_3^{x'\tau}(y')+\int \diff x'Q_2^{y'\tau}(x') \right] . \label{a:SU(3)Qprime}
\end{align}
It also holds $Q_\alpha^{x'\tau},\; Q_\alpha^{y'\tau}\in\mathbb{Z}$ because they are topological charges in the $x'\tau$- and $y'\tau$-plane. At this time,  we return the integral to the lattice sum: 
\begin{equation}  \label{a:SU(3)resum}
\frac{1}{\sqrt{3}a}\int \diff y' \longmapsto \sum_k \:,\quad \frac{1}{3a}\int \diff x' \longmapsto \sum_j \: . 
\end{equation}
$\vec{\phi}_\alpha$ changes continuously in the $a\to0$ limit, but $Q_\alpha^{\mu'\tau}\;(\mu'=x'\text{ or } y')$ is only integer-valued, so $Q_\alpha^{\mu'\tau}$ is constant. 

We put the system on the large torus, $T^2$, with the periodic boundary condition. 
In order to take the periodic boundary condition without any frustration, the number of sites along $x,y$ directions must be both in multiples of $3$. 
This requires that $k$ takes the value of $1$ to $3M \; (M\in\mathbb{N})$ while $j$ takes $1$ to arbitrary natural number $M'$. As a result, 
\begin{align}
S_\mathrm{add} &= -2\pi \im \, p\left[ \frac{1}{3}\sum_{k=1}^{3M}Q_1^{x'\tau}+\frac{2}{3}\sum_{k=1}^{3M}Q_3^{x'\tau}+\sum_{j=1}^{M'}Q_2^{y'\tau} \right] \nonumber \\ 
&= -2\pi \im\,p\left[ \frac{1}{3}\cdot Q_1^{x'\tau}\cdot3M+\frac{2}{3}\cdot Q_3^{x'\tau}\cdot3M+Q_2^{y'\tau}\cdot M' \right] \nonumber \\ 
&= -2\pi \im\, p\left[ (Q_1^{x'\tau}+2Q_3^{x'\tau})\cdot M+Q_2^{y'\tau}\cdot M' \right] \qquad \in2\pi \im \mathbb{Z} \;.   \label{a:SU(3)vanish}
\end{align}
Hence, the topological term does not contribute to the partition function $Z$.

\section{Uniqueness of the effective Lagrangian}  \label{sec:B}

In Appendix~\ref{sec:B}, we prove that the nonlinear sigma model~\eqref{SU(3)effaction} is the unique possibility, which is constructed by the $SU(3)$ matrix-valued field, $U(x,y,\tau)=(\vec{\phi}_{\alpha}(x,y,\tau)^{\mathrm{T}})$, within the second-order spacetime derivatives and satisfies the symmetry requirement.
The relevant symmetries are given by the $U(1)^2$ gauge invariance, $PSU(3)$ spin rotational symmetry, the $p6m$ lattice symmetry, and the time-reversal symmetry, as discussed in Sec.~\ref{sec:symmetry}. 
Repeated subscripts without the symbol $\sum$ are \textit{not} summed in this section. 

To begin with, let us exclude the first-order derivative terms. 
Because of the $PSU(3)$ invariance, it must take the form of $\vec{\phi}_{\alpha}^*\cdot \partial_\mu \vec{\phi}_\beta$. 
For $\alpha\not=\beta$, it is not $U(1)\times U(1)$ gauge invariant, so we only need to consider $\vec{\phi}_{\alpha}^*\cdot \partial_\mu \vec{\phi}_\alpha$. 
They are also not manifestly $U(1)$ gauge invariant either, but the change is given by the total derivative. 
As the one-unit lattice translation changes $\vec{\phi}_{\alpha}\mapsto \vec{\phi}_{\alpha+1}$, we must consider the invariant combination, but it automatically vanishes because of the $SU(3)$ constraint:
\begin{equation} \label{b:drop4}
\sum_\alpha\vec{\phi}_\alpha^\ast\cdot\partial_\mu\vec{\phi}_\alpha
= \mathrm{tr}\left[U^\dagger \partial_\mu U\right] =0 \;. 
\end{equation}
Here, we used the fact that $U^\dagger \partial_\mu U$ is an element of the Lie algebra, $\mathfrak{su}(3)$, which is an anti-Hermitian, traceless matrix. 
One may have thought that we can more easily exclude the first-order derivative terms using the time-reversal, and $x$-, $y$-axes reflections, or some subgroup of $\mathbb{Z}_6$ rotation. 
For the spatial derivative, $\partial_{x,y}$, this is indeed true, but we cannot exclude the time derivative, $\partial_{\tau}$ using this logic. 
This is because the time-reversal operation flips the ``real'' time $t$ with the complex conjugation of the $c$-numbers: $\vec{\phi}_\alpha^\ast\cdot\partial_\tau\vec{\phi}_\alpha$ is the Wick rotation of $\im\,\vec{\phi}_\alpha^\ast\cdot\partial_t\vec{\phi}_\alpha$, and it is invariant under $t\mapsto -t$ combined with the complex conjugation, $\im\mapsto -\im$. 
Therefore, it seems that the use of the one-unit lattice translation is mandatory to exclude the first-order derivative term of the imaginary time. 

As we have excluded the first-order derivative terms, we concentrate on the second-order derivative terms in the following discussions. 
Because of $PSU(3)$ symmetry~\eqref{SU(3)symS}, 
we must construct the $SU(3)$ scalars using the vectors $\vec{\phi}_\alpha$ and their conjugates, and there are two possible ways for the construction. 
The first one is to use the scalar product between $\vec{\phi}_\alpha$ and $\vec{\phi}_\beta^*$, which can be given as 
\begin{equation}  \label{b:secondderiv}
\partial_\mu\vec{\phi}_\alpha\cdot\partial_\nu\vec{\phi}_\beta^\ast \;,\quad 
(\partial_\mu\vec{\phi}_\alpha\cdot\vec{\phi}_\beta^\ast)(\partial_\nu\vec{\phi}_\gamma\cdot\vec{\phi}_\delta^\ast) \;.
\end{equation}
The second one is to use the determinant as the trivial representation of $SU(3)$, 
\begin{equation} \label{b:secondderive_det}
    \vec{\phi}_{\alpha}\cdot (\partial_\mu \vec{\phi}_\beta\times \partial_\nu\vec{\phi}_{\gamma}) \;. 
\end{equation}
Here, $\mu,\nu=x,y,\tau$ and $\alpha,\beta,\gamma,\delta=1, 2, 3$. 
We do not obtain anything new by multiplying \eqref{b:secondderiv} and \eqref{b:secondderive_det} with non-derivative fields because $\vec{\phi}_{\alpha}$ satisfies the orthonormality constraint, \eqref{orthonomal}; it is just 1 or 0. Zeroth-derivative terms are also meaningless for the same reason. 

It is sufficient to discuss the form of \eqref{b:secondderiv} in the following, because \eqref{b:secondderive_det} can be reduced to the terms in \eqref{b:secondderiv}. 
To see this, we note that the $U(1)^2$ gauge invariance of \eqref{b:secondderive_det} requires that $\alpha,\beta,\gamma$ must be some permutation of $1,2,3$, so the typical example is given by $\vec{\phi}_{1}\cdot (\partial_\mu \vec{\phi}_2\times \partial_\nu \vec{\phi}_{3})$.  
Because of the $SU(3)$ property, we can express $\vec{\phi}_3$ as $\vec{\phi}_3=\vec{\phi}_1^*\times \vec{\phi}_2^*$. 
As a result, we find $\vec{\phi}_{1}\cdot (\partial_\mu \vec{\phi}_2\times \partial_\nu \vec{\phi}_{3})=\vec{\phi}_{1}\cdot (\partial_\mu \vec{\phi}_2\times \partial_\nu (\vec{\phi}_1^*\times \vec{\phi}_2^*))$, and the simple vector calculus shows that this can be expressed in terms of \eqref{b:secondderiv}. 

For the $U(1)^2$ gauge invariance of \eqref{b:secondderiv}, the phase factor $\mathrm{e}^{i\vartheta_\alpha}$ arisen from $\vec{\phi}_\alpha$ must be compensated by $\vec{\phi}_\alpha^\ast$. 
The indices in \eqref{b:secondderiv} can be restricted as 
\begin{equation}  \label{b:grobalgauge}
(\partial_\mu\vec{\phi}_\alpha\cdot\partial_\nu\vec{\phi}_\alpha^\ast) \;, \quad
(\partial_\mu\vec{\phi}_\alpha\cdot\vec{\phi}_\alpha^\ast)(\partial_\nu\vec{\phi}_\beta\cdot\vec{\phi}_\beta^\ast) \;, \quad
(\partial_\mu\vec{\phi}_\alpha\cdot\vec{\phi}_\beta^\ast)(\partial_\nu\vec{\phi}_\beta\cdot\vec{\phi}_\alpha^\ast) \;. 
\end{equation}
The $U(1)^2$ gauge invariance~\eqref{SU(3)symG} gives a stronger constraint, and we perform the explicit calculation of local gauge transformation law of \eqref{b:grobalgauge}:
% \footnote{Actually, the gauge invariance of eqs.~\eqref{b:gauge1}--\eqref{b:gauge6} holds without summation of $\alpha$. }. 
\begin{align}
\partial_\mu\vec{\phi}'_\alpha\cdot\partial_\nu\vec{\phi}_\alpha^{\prime\ast}
&=\partial_\mu\vec{\phi}_\alpha\cdot\partial_\nu\vec{\phi}_\alpha^{\ast}
+\im\, (\partial_\mu\vartheta_\alpha)\vec{\phi}_\alpha\cdot\partial_\nu\vec{\phi}_\alpha^\ast+\im\, (\partial_\nu\vartheta_\alpha)\vec{\phi}_\alpha\cdot\partial_\mu\vec{\phi}_\alpha^\ast+(\partial_\mu\vartheta_\alpha)(\partial_\nu\vartheta_\alpha) \;,  \label{b:gauge1}\\
(\partial_\mu\vec{\phi}'_\alpha\cdot\vec{\phi}_\alpha^{\prime\ast})(\partial_\nu\vec{\phi}'_\beta\cdot\vec{\phi}_\beta^{\prime\ast})
&=(\partial_\mu\vec{\phi}_\alpha\cdot\vec{\phi}_\alpha^\ast)(\partial_\nu\vec{\phi}_\beta\cdot\vec{\phi}_\beta^\ast)
-\im\, (\partial_\mu\vartheta_\alpha)\vec{\phi}_\beta\cdot\partial_\nu\vec{\phi}_\beta^\ast
-\im\,(\partial_\nu\vartheta_\beta)\vec{\phi}_\alpha\cdot\partial_\mu\vec{\phi}_\alpha^\ast
-(\partial_\mu\vartheta_\alpha)(\partial_\nu\vartheta_\beta) \;,  
\label{b:gauge2}\\
(\partial_\mu\vec{\phi}'_\alpha\cdot\vec{\phi}_{\beta}^{\prime\ast})(\partial_\nu\vec{\phi}'_{\beta}\cdot\vec{\phi}_\alpha^{\prime\ast})
&=(\partial_\mu\vec{\phi}_\alpha\cdot\vec{\phi}_{\beta}^\ast)(\partial_\nu\vec{\phi}_{\beta}\cdot\vec{\phi}_\alpha^\ast)\;, \quad (\alpha\not=\beta)\;.  \label{b:gauge4}
\end{align}
In \eqref{b:gauge2}, we can take both $\alpha=\beta$ and $\alpha\not=\beta$, while we have to take $\alpha\not=\beta$ in \eqref{b:gauge4}. 
To obtain the expressions on the right-hand-side, we use several identities following from the orthonormality. 
The last one~\eqref{b:gauge4} is apparently gauge invariant, but others are not, and we must take specific combinations to achieve the gauge invariance. 
There are three ways to satisfy it: 
\begin{itemize}
\item[(i)]  
Since \eqref{b:gauge1} and \eqref{b:gauge2} have the common extra term with the opposite sign when we set $\alpha=\beta$ in \eqref{b:gauge2}, the following combination is gauge invariant: 
\begin{equation}  \label{b:gauge5}
(\partial_\mu\vec{\phi}_\alpha\cdot\partial_\nu\vec{\phi}_\alpha^\ast)
+(\partial_\mu\vec{\phi}_\alpha\cdot\vec{\phi}_\alpha^\ast)(\partial_\nu\vec{\phi}_\alpha\cdot\vec{\phi}_\alpha^\ast) \;. 
\end{equation}
\item[(ii)] 
Since the extra terms in \eqref{b:gauge1} are symmetric about $\mu$ and $\nu$, so the asymmetrized combination,
\begin{equation}  \label{b:gauge6}
(\partial_\mu\vec{\phi}_\alpha\cdot\partial_\nu\vec{\phi}_\alpha^\ast)-(\partial_\nu\vec{\phi}_\alpha\cdot\partial_\mu\vec{\phi}_\alpha^\ast)\;,  \quad(\mu\neq\nu)\;,
\end{equation}
is gauge invariant. 
Although the same logic applies to \eqref{b:gauge2} with $\alpha=\beta$, it is symmetric by itself, so the asymmetrization gives $0$ and we do not get the meaningful gauge-invariant term. 
\item[(iii)] As the gauge parameter is subject to the constraint, $\sum_\alpha\vartheta_\alpha=0$, 
we can construct the gauge-invariant term when the extra term has the form of $(\sum_\beta\vartheta_\beta)\vec{\phi}_\alpha\cdot\partial_\nu\vec{\phi}_\alpha^\ast=0$. 
To find such a combination, we can sum over $\alpha,\beta$ independently in \eqref{b:gauge2}, but it gives 
\begin{align}
\left( \sum_\alpha\vec{\phi}_\alpha^\ast\cdot\partial_\mu\vec{\phi}_\alpha \right)\left( \sum_\beta\vec{\phi}_\beta^\ast\cdot\partial_\nu\vec{\phi}_\beta \right) =0 \;,   \label{b:gauge8}
\end{align}
because of the identity \eqref{b:drop4}. We do not obtain new terms for the possible effective Lagrangian. 
\end{itemize}
As a result, we have found the gauge invariant terms, \eqref{b:gauge4}--\eqref{b:gauge6}. 
When the indices $\mu\nu$ in \eqref{b:gauge4} is symmetrized, the same manipulation in \eqref{a:SU(3)Uphikin} shows that it is identical with \eqref{b:gauge5}. 
Therefore, the spacetime indices in \eqref{b:gauge4} should be anti-symmetrized to obtain the independent term from \eqref{b:gauge5}. 
Physically, \eqref{b:gauge5} is a generalization of the kinetic term, \eqref{b:gauge6} corresponds to the $2$d topological term (or the $U(1)$ field strength), $Q_{\alpha}^{\mu\nu}$, and the anti-symmetrization of \eqref{b:gauge4} corresponds to the torsion term of the sigma model, which has appeared in the $2$d sigma model of Ref.~\cite{Lajko:2017wif} and is called as the $\lambda$ term. 

So far, we have only used the $U(1)^2$ gauge invariance and the $PSU(3)$ spin rotational symmetry, 
so we constrain the possible terms by using the lattice symmetry in the following. 
As a consequence of the one-unit lattice translation, the subscripts $\alpha=1,2,3$ must enter equally in Lagrangian for $\mathbb{Z}_3$ cyclic symmetry~\eqref{SU(3)symZ} to hold. 
The allowed distinct terms can be obtained from \eqref{b:gauge4} and \eqref{b:gauge5}:
\begin{align}
    &\sum_{\alpha}\sum_{\nu,\rho}\ve_{\mu\nu\rho}(\partial_\nu \vec{\phi}_\alpha\cdot \vec{\phi}^*_{\alpha+1})(\partial_\rho \vec{\phi}_{\alpha+1}\cdot \vec{\phi}^*_\alpha) \;, \\
    &\sum_{\alpha}\left[(\partial_\mu\vec{\phi}_\alpha\cdot\partial_\nu\vec{\phi}_\alpha^\ast)
+(\partial_\mu\vec{\phi}_\alpha\cdot\vec{\phi}_\alpha^\ast)(\partial_\nu\vec{\phi}_\alpha\cdot\vec{\phi}_\alpha^\ast)\right] . 
\end{align}
We note that the summation over $\alpha$ of \eqref{b:gauge6} vanishes, since $\tr[\diff U\wedge \diff U^\dagger]=\diff\{\tr[U\diff U^\dagger]\}=0$ as a consequence of \eqref{b:drop4}. 

We then consider the spatial rotational symmetry. We note that $\mathbb{Z}_6$ rotation around a site of $\alpha=1$ can be decomposed into the $\mathbb{Z}_3$ rotation and the $\mathbb{Z}_2$ rotation. 
Especially, we would like to emphasize that the $\mathbb{Z}_3$ rotation does not change the indices $\alpha$, while the $\mathbb{Z}_2$ rotation exchanges $\alpha=2$ and $3$. 
To have the $\mathbb{Z}_3$ rotational invariance with two spacetime indices, we must construct the scalar under the spatial $SO(2)$ rotation.
We therefore obtain the following two terms as the kinetic term from \eqref{b:gauge5}:
\begin{align}
\label{b:refrot1}
&\sum_\alpha\left[(\partial_\tau\vec{\phi}_\alpha\cdot\partial_\tau\vec{\phi}_\alpha^\ast)
+(\partial_\tau\vec{\phi}_\alpha\cdot\vec{\phi}_\alpha^\ast)(\partial_\tau\vec{\phi}_\alpha\cdot\vec{\phi}_\alpha^\ast)\right] ,  \\
\label{b:refrot2}
&\sum_\alpha\sum_{\mu=x,y}\left[(\partial_\mu\vec{\phi}_\alpha\cdot\partial_\mu\vec{\phi}_\alpha^\ast)
+(\partial_\mu\vec{\phi}_\alpha\cdot\vec{\phi}_\alpha^\ast)(\partial_\mu\vec{\phi}_\alpha\cdot\vec{\phi}_\alpha^\ast)\right] ,  
\end{align}
and, from \eqref{b:gauge4}, we obtain 
\begin{align}\label{b:refrot3}
    \sum_{\alpha}\ve_{\tau\nu\rho}(\partial_\nu \vec{\phi}_\alpha\cdot \vec{\phi}^*_{\alpha+1})(\partial_\rho \vec{\phi}_{\alpha+1}\cdot \vec{\phi}^*_\alpha)
\end{align}
The last term~\eqref{b:refrot3}, however, cannot be simultaneously invariant under the $\mathbb{Z}_2$ rotation and the $x$-axis reflection. 
Since \eqref{b:refrot3} is already a scalar under the $SO(2)$ spatial rotation, the $\pi$ rotation requires the invariance under the exchange of $\alpha=2$ and $3$, so that the $x$-axis reflection becomes the same with $x\mapsto -x$ without affecting the indices $\alpha$. 
As \eqref{b:refrot3} contains the single derivative $\p_x$, the $x$-axis reflection flips its sign, so it is excluded from the list of possible terms of Lagrangian. 
We can easily check that \eqref{b:refrot1} and \eqref{b:refrot2} are invariant under the whole set of the lattice symmetry and the time-reversal symmetry. 
After the rescaling of the imaginary time, we find that \eqref{SU(3)effaction} is the unique possible term of the effective Lagrangian within the second-order spacetime derivatives, and this completes the proof. 

Let us comment on some subtleties related to the topological-like terms~\eqref{b:gauge6} before closing this section. Since they may have another mechanism to realize the symmetries as a result of their quantization, we have to be more careful on the discussion related to the lattice symmetry. 
For example, the $\mathbb{Z}_3$ permutation symmetry can be realized without summing over $\alpha$ of \eqref{b:gauge6} by a suitable quantization of the coefficient, such as ${2\pi \im\over 3}(Q_2^{\mu\nu}-Q_3^{\mu\nu})$, due to the integral nature of topological charges: Under the $\mathbb{Z}_3$ permutation, ${2\pi\im\over 3}(Q_2^{\mu\nu}-Q_3^{\mu\nu})\mapsto {2\pi \im \over 3}(Q_3^{\mu\nu}-Q_1^{\mu\nu})={2\pi\im\over 3}(Q_2^{\mu\nu}-Q_3^{\mu\nu})+2\pi \im Q_3^{\mu\nu}$ thanks to $Q_1+Q_2+Q_3=0$.
This subtlety requires the extra care, as we have done in Appendix~\ref{ssec:apSU(3)top}.

%%%%%%%%%%%%%%%%%%%%%%%%%%%%%%%%%%%%%%%%%%%%%%%%%%%%%%%%%%%%%%%%%%%%%%%%%%%%%%%%%%%%%%%%%%%%%%%%%%%%
\section{Abelian duality and monopoles}\label{sec:3d_Abelian_duality}

Compact $U(1)$ gauge theory can be dualized to the periodic scalar theory in $3$d. 
Using this $3$d Abelian duality, we can analyze the monopole-instanton gas in the semiclassical way~\cite{Polyakov:1976fu}. 
In this appendix, we apply the Abelian duality to the $U(1)^2$ gauge theory with the following kinetic term,
\begin{equation}
    \mathcal{L}_{\mathrm{kin}}={1\over 2e^2}\left(|\diff a_1|^2+|\diff a_2|^2+|\diff a_1+\diff a_2|^2\right) \;. 
    \label{eq:compact_gauge}
\end{equation}
Here, $a_\alpha$ are dynamical $U(1)$ gauge fields with $\alpha=1,2$, and we denote $|\diff a_\alpha|^2=\diff a_\alpha \wedge \star \diff a_\alpha$. 
There is a mixed kinetic term, $\diff a_1\wedge \star \diff a_2$ in order to satisfy the $S_3$ permutation symmetry as we have seen in the $SU(3)/U(1)^2$ sigma model. 
This $S_3$ symmetry becomes more obvious by introducing $a_3=-(a_1+a_2)$ as we have done in the main text, then the action of $s\in S_3$ is given by  $a_\alpha\mapsto a_{s(\alpha)}$. 
A compact $U(1)^{N-1}$ gauge theory with $S_N$ gauge symmetry has been analyzed in Ref.~\cite{Nguyen:2021yld}, but we here leave $S_N$ as a global symmetry.

To obtain the dual theory, it is convenient to introduce the auxiliary two-form fields $b_{1}$ and $b_{2}$, and replace $\diff a_\alpha$ by $(\diff a_{\alpha}-b_{\alpha})$. 
This operation changes the theory, but we can recover the original theory by introducing topological couplings with $2\pi$-periodic scalar fields $\sigma_1$ and $\sigma_2$:
\begin{align}
    \mathcal{L}'_{\mathrm{kin}}
    &={1\over 2e^2}\left(|\diff a_1-b_1|^2+|\diff a_2-b_2|^2+|(\diff a_1-b_1)+(\diff a_2-b_2)|^2\right) \nonumber\\
    &\quad +\im{1\over 2\pi}\left(b_1\wedge \diff \sigma_1 + b_2\wedge \diff \sigma_2\right) \;. 
    \label{eq:compact_gauge_dual}
\end{align}
Let us explain why the theory defined by \eqref{eq:compact_gauge_dual} is equivalent to \eqref{eq:compact_gauge}. 
We first integrate out the $2\pi$-periodic scalar fields $\sigma_1,\sigma_2$. 
By using the equation of motion for $\sigma_\alpha$, we obtain 
\begin{equation}
    \diff b_\alpha=0 \;. 
    \label{eq:eom_sigma}
\end{equation}
Therefore, at least locally, we can rewrite the two-form field $b_\alpha$ with some one-form field $c_\alpha$ as 
\begin{equation}
    b_\alpha=\diff c_{\alpha} \;. 
    \label{eq:solve_eom_sigma}
\end{equation}
Moreover, because of the $2\pi$ periodicity of $\sigma_\alpha$, we can find that 
\begin{equation}
    \int_{M_2}\diff c_\alpha \in 2\pi \mathbb{Z} \;, 
\end{equation}
where $M_2$ is an arbitrary closed $2$-manifold in $3$-dimensional spacetime. 
This means that $c_\alpha$ can be regarded as a $U(1)$ gauge field, and $b_\alpha$ is its field strength. 
By replacing $a_\alpha-c_\alpha$ by $a_\alpha$, we recover the original Lagrangian~\eqref{eq:compact_gauge}. 

Next, let us perform the path integral in the opposite way. 
Instead of integrating out $\sigma_\alpha$, we perform the path integral of $a_\alpha$ and $b_{\alpha}$. 
For this purpose, we consider the equation of motion in terms of $b_{\alpha}$,
\begin{align}
    2b_1+b_2&=2\diff a_1+\diff a_2-\star \left(\im {e^2\over 2\pi}\diff \sigma_1\right), \\
    b_1+2b_2&=\diff a_1+2\diff a_2 -\star \left(\im{e^2\over 2\pi}\diff \sigma_2\right). 
\end{align}
This can be solved as 
\begin{align}
    b_1&=\diff a_1-\star\, \im{e^2\over 2\pi}\left({2\over 3}\diff \sigma_1-{1\over 3}\diff \sigma_2\right),\\
    b_2&=\diff a_2-\star\, \im {e^2\over 2\pi}\left({2\over 3}\diff \sigma_2-{1\over 3}\diff \sigma_1\right). 
\end{align}
Substituting this expression into \eqref{eq:compact_gauge_dual}, we obtain the kinetic term of the dual Lagrangian,
\begin{equation}
    \mathcal{L}_{\mathrm{kin}}^{\mathrm{dual}}={e^2\over 24\pi^2}\left(|\diff \sigma_1|^2+|\diff \sigma_2|^2+|\diff \sigma_1-\diff \sigma_2|^2\right).
    \label{eq:dual_kinetic}
\end{equation}
In deriving this expression, we note that 
\begin{equation}
    \int_{M_3}{1\over 2\pi}\diff a_\alpha\wedge \diff \sigma_\beta\in 2\pi \mathbb{Z} ;\, 
\end{equation}
for any closed $3$-manifolds because of the Dirac quantization of $a_\alpha$ and the $2\pi$-periodicity of $\sigma_\alpha$. 

Let us check how the $S_3$ permutation symmetry acts on the dual theory~\eqref{eq:dual_kinetic}. 
The $\mathbb{Z}_3$ cyclic permutation, $a_\alpha\mapsto a_{\alpha+1\,}$, is given by 
\begin{equation}
    \begin{pmatrix}
    \diff \sigma_1\\
    \diff \sigma_2
    \end{pmatrix}\mapsto
    \begin{pmatrix}
    -1 & 1\\
    -1 & 0
    \end{pmatrix}
    \begin{pmatrix}
    \diff \sigma_1\\
    \diff \sigma_2
    \end{pmatrix}. 
    \label{eq:Z3permutation_dual}
\end{equation}
We note that this transformation rule is same with that of $A^{\mathrm{top}}_{1,2}$ given in \eqref{eq:monopole_gauge_field_permutation}. 
The $\mathbb{Z}_2$ permutation, $a_1\mapsto a_1$ and  $a_2\mapsto a_3=-(a_1+a_2)$, is given by 
\begin{equation}
    \begin{pmatrix}
    \diff \sigma_1\\
    \diff \sigma_2
    \end{pmatrix}\mapsto
    \begin{pmatrix}
    1 & -1\\
    0 & -1
    \end{pmatrix}
    \begin{pmatrix}
    \diff \sigma_1\\
    \diff \sigma_2
    \end{pmatrix}. 
    \label{eq:Z2permutation_dual}
\end{equation}
These transformation properties can be obtained by requiring the invariance of \eqref{eq:compact_gauge_dual}. 
At this stage, we can only identify the transformation law for the derivatives, $\diff \sigma_\alpha$. 
When identifying the transformation property of the fields themselves, there is an ambiguity of additive constants, and we have to study the property of monopole operators to fix it. 

Because of the $2\pi$ periodicity of $\sigma_\alpha$, the local operator should take the form of 
\begin{equation}
    \exp(\im \sigma_\alpha(x)) \;. 
\end{equation}
This corresponds to the monopole operator in the language of $U(1)$ gauge theory. 
This can be seen by noting that the insertion of $\rme^{\im \sigma_\alpha(x_*)}$ at $x=x_*$ modifies  the equation of motion~\eqref{eq:eom_sigma} as 
\begin{equation}
    {1\over 2\pi}\diff b_\alpha=\delta^{(3)}(x-x_*)\diff\tau\wedge \diff x\wedge \diff y \;. 
\end{equation}
Let us consider a small sphere, $S^2_*$, around $x_*$, then the Stokes' theorem gives 
\begin{equation}
    \int_{S^2} b_\alpha=2\pi \;. 
\end{equation}
By removing the point $x_*$ from the spacetime, we can introduce the $U(1)$ gauge field $c_\alpha$ by \eqref{eq:solve_eom_sigma}, with the above boundary condition around $x$. 
This is nothing but the $U(1)$ gauge field with the Dirac monopole at $x$. 
In order to recover the original gauge theory \eqref{eq:compact_gauge}, we must regard $a_\alpha-c_\alpha$ as $a_\alpha$, and then $a_\alpha$ inherits the monopole singularity of $c_\alpha$. 
Therefore, $\rme^{\im \sigma_1}$ introduces the monopole with charge $(Q_1, Q_2, Q_3)=(-1, 0 , 1)$. The same is true for $\rme^{\im \sigma_2}$, and it has the charge $(Q_1,Q_2,Q_3)=(0,-1,1)$. 

If we naively think that $S_3$ permutation acts on $\sigma_\alpha$ by \eqref{eq:Z3permutation_dual} and \eqref{eq:Z2permutation_dual} without additive constants even after removing derivatives, then the following monopole potential can be added to the Lagrangian:
\begin{equation}
    V_{\mathrm{eff}}=-\left(\cos(\sigma_1)+\cos(\sigma_1-\sigma_2)+\cos(\sigma_2)\right) \;. 
\end{equation}
Such a monopole potential is expected to be generated under the renormalization-group flow unless there is a symmetry forbidding it. 
We here put the overall negative sign, as it typically appears when the dilute gas approximation is valid.  
In this case, the system becomes gapped with the unique ground state, $\sigma_1=\sigma_2=0 \bmod 2\pi$.

\bibliographystyle{utphys}%\typeout{}
\bibliography{./antiferroBIB,./QFT,./refs}

\end{document}